\documentclass[11pt]{article}
\usepackage{color,amsfonts,amsmath,amssymb,mathrsfs,verbatim,graphicx}

\def\bdx{\mbox{\boldmath $\delta x$}}
\def\bxn{\mbox{\boldmath $\delta x$}_n}
\def\bxnf{\mbox{\boldmath $\delta x$}_{n-4}}

\def\rrr{{\mathbb{R}}}
\def\ccc{{\mathbb{C}}}
\def\hhh{{\mathbb{H}}}
\def\ooo{{\mathbb{O}}}

\def\hkc{\mbox{h}_k\ccc}
\def\htwc{\mbox{h}_2\ccc}
\def\hthc{\mbox{h}_3\ccc}
\def\htho{\mbox{h}_3\ooo}
\def\fhtho{F(\mbox{h}_3\ooo)}
\def\hpc{\mbox{h}_p\ccc}

\def\sltc{\mbox{SL}(2,\ccc)}
\def\slthc{\mbox{SL}(3,\ccc)}
\def\sltho{\mbox{SL}(3,\ooo)}
\def\slpc{\mbox{SL}(p,\ccc)}
\def\slptc{\mbox{SL}(k,\ccc)_{\! D}}
\def\sukd{\mbox{SU}(k)_{\! D}}
\def\uod{\mbox{U}(1)_{\! D}}

\def\gt{\mbox{G}_2}
\def\ff{\mbox{F}_4}
\def\esi{\mbox{E}_6}
\def\ese{\mbox{E}_7}
\def\ee{\mbox{E}_8}
\def\esig{\mbox{E}_{6(-26)}}
\def\eseg{\mbox{E}_{7(-25)}}
\def\eeg{\mbox{E}_{8(-24)}}

\def\bone{\mbox{\boldmath $1$}}
\def\bthr{\mbox{\boldmath $3$}}

\def\soe{\mbox{SO}(8)}
\def\soot{\mbox{SO}^+(1,3)}

\def\sompmm{\mbox{SO}(m_+,m_-)_{\! D}}

\def\somdp{\mbox{SO}(m_+)_{\! D}}

\def\bv{\mbox{\boldmath $v$}}

\def\sutw{\mbox{SU}(2)}
\def\suth{\mbox{SU}(3)}
\def\sufo{\mbox{SU}(4)}
\def\uo{\mbox{U}(1)}
\def\hG{\hat{G}}

\def\Lam{\mbox{\tiny $\Lambda$}}

\def\pal{\partial}

\def\lag{\mathcal{L}}
\def\pha{\phantom}
\def\cabg{c^{\alpha}_{\pha{o}\beta\gamma}}

\def\Mxl{\mbox{\tiny M}}
\def\Yml{\mbox{\tiny YM}}

\def\Amin{A_{\mbox{-}}}
\def\Apls{A_{\mbox{\tiny{+}}}}
\def\Apm{A_{\mbox{\tiny{$\pm$}}}}

\def\pAp{p_{\mbox{\tiny$\! A_{\;\!\!+}$}}}
\def\pAm{p_{\mbox{\tiny$\! A$}_{\mbox{-}}}}

\def\rAp{\rho_{\mbox{\tiny$\! A_{\;\!\!+}$}}}
\def\rAm{\rho_{\mbox{\tiny$\! A$}_{\mbox{-}}}}

\def\wAp{w_{\mbox{\tiny$\! A_{\;\!\!+}$}}}
\def\wAm{w_{\mbox{\tiny$\! A$}_{\mbox{-}}}}

\def\pApm{p_{\mbox{\tiny$\! A_{\;\!\!\pm}$}}}
\def\rApm{\rho_{\mbox{\tiny$\! A_{\;\!\!\pm}$}}}
\def\wApm{w_{\mbox{\tiny$\! A_{\;\!\!\pm}$}}}

\def\pAMp{p_{\mbox{\tiny$\! AM_{\;\!\!+}$}}}
\def\rAMp{\rho_{\mbox{\tiny$\! AM_{\;\!\!+}$}}}
\def\wAMp{w_{\mbox{\tiny$\! AM_{\;\!\!+}$}}}

\def\pvac{p_{\mbox{\tiny$V$}}}
\def\rvac{\rho_{\mbox{\tiny$V$}}}
\def\wvac{w_{\mbox{\tiny$V$}}}

\def\bphi{\mbox{\boldmath $\phi$}}
\def\bpsi{\mbox{\boldmath $\psi$}}
\def\Amax{A^{\mbox{\raisebox{+3pt}{\tiny\boldmath$\!\!\!\circ$}}}}

\def\bphip{\mbox{\boldmath $\phi$}_{\mbox{\tiny{\!+}}}}
\def\bpsip{\mbox{\boldmath $\psi$}_{\mbox{\tiny{\!+}}}}
\def\Amaxp{A^{\mbox{\raisebox{+3pt}{\tiny\boldmath$\!\!\!\circ$}}}_{\mbox{\tiny{+}}}}

\def\Mpls{\bphip,\bpsip,\Amaxp}
\def\Mplu{M_{\mbox{\tiny{+}}}}

\def\setb{\setlength{\baselineskip}{0.625\baselineskip}}

\begin{document} 

{\setlength{\baselineskip}{0.625\baselineskip}

\begin{center}
   
           {\LARGE {\bf Generalised Local Extra Dimensions as a Basis}}  
                 
              \vspace{10pt}          
              
            {\LARGE {\bf for the Elementary Structure of Matter}}  

   \vspace{25pt}
 
 \mbox {{\Large David J. Jackson}} \\ 

    \vspace{12pt}  
 
  {david.jackson.th@gmail.com}  \\

   \vspace{12pt}
 
 {August 25, 2022}

 \vspace{30pt}

{\bf  Abstract}

 
\end{center}

 A central aim of theoretical physics is to account for the structure of matter at the most elementary level as underlying the Standard Model of particle physics, and ideally also as a basis for a substantial dark sector, as distributed in 4-dimensional spacetime.
 A broad class of theories augment the global \mbox{4-dimensional} spacetime arena itself to a 
    higher-dimensional spacetime structure, with the properties of matter then deriving from the properties of the extra spatial dimensions.
     We motivate an alternative approach in which we begin  with the local structure of \mbox{4-dimensional} spacetime and augment the local form for a proper time interval. The symmetries and properties of the residual components, over the local 4-dimensional spacetime form, are found to exhibit appropriate features to account for the visible Standard Model sector while in parallel direct connections can also be identified with dark sector models.

\vspace{10pt}

\tableofcontents



\pagebreak

\section{Introduction}
\label{bas1}

    The familiar yet esoteric properties of the Standard Model of particle physics, with further theoretical understanding desired in particular in the Higgs and neutrino sectors, together with the need to account for a significant dark sector as evident on galactic and cosmological scales, provide a clear and extensive target for any proposed unification scheme.
    More generally, the origin of the elementary particle multiplet structure as deduced from high energy physics experiments, and the source of the accelerating expansion of the universe together with the nature of a dark matter component as deduced from cosmological observations, rank amongst the most pressing and outstanding questions to be addressed in fundamental physics.

      Amongst unification schemes that employ a generalised spacetime structure the addition of postulated 
      extra spatial dimensions to the familiar global 4-dimensional spacetime has become the dominant paradigm. 
         However, no compelling success in accounting for the Standard Model or other observed phenomena has been achieved in this manner and there remains no evidence that additional dimensions of space actually exist (\cite{PDG22} section~85). 
                In this paper we describe an alternative means of augmenting 4-dimensional spacetime, 
            with the  emphasis upon the local structure, through a direct generalisation of the form for a local proper time interval. We argue that this new approach provides a far more suitable basis for the structure of matter and for a unified theory generally.
            
            In the following section we first motivate the basis for this theory.
           After reviewing the connections that have been established with the Standard Model in section~\ref{bas3}   we describe in section~\ref{bas4} how a further structure of matter can be identified for this theory that has the potential to account for the phenomena of a dark sector. 
                While being conceptually well-motivated in itself  the ability of this new approach     
                         to accommodate the principal empirical features of \textit{both} the visible and dark sectors, and 
                  also point forward to areas where new physics may be uncovered, makes a strong case for this alternative to the extra spatial dimension paradigm. 
                   The suitability of the new theory as an overall unification scheme, with the potential to also  provide an underlying basis for the principle of extremal action, will be further  assessed  in section~\ref{bas5}, before we conclude in the final section.


\section{Generalised Proper Time as the Source of Matter}
\label{bas2}

     For any theory of the elementary microscopic structure of matter, such as relating to the Standard Model of particle physics,  we are interested in the properties of local symmetries and local particle interactions. This motivates beginning with the 
     local structure of 4-dimensional spacetime as the basis for identifying the properties of matter from the components of an augmented structure. A direct means of augmenting the local spacetime metric structure is provided by the form of a proper time interval $\delta s$ expressed in a local inertial reference frame (\cite{Gener} section~2,  as reviewed here). Compared with the familiar approach involving extra dimensions of space, the change of focus in this starting point and contrasting mechanism for augmentation is depicted in figure~\ref{contra}.     
     
     \pagebreak
     
\begin{figure}[htbp]  
\centering
\leavevmode
\includegraphics[width=14.4cm]{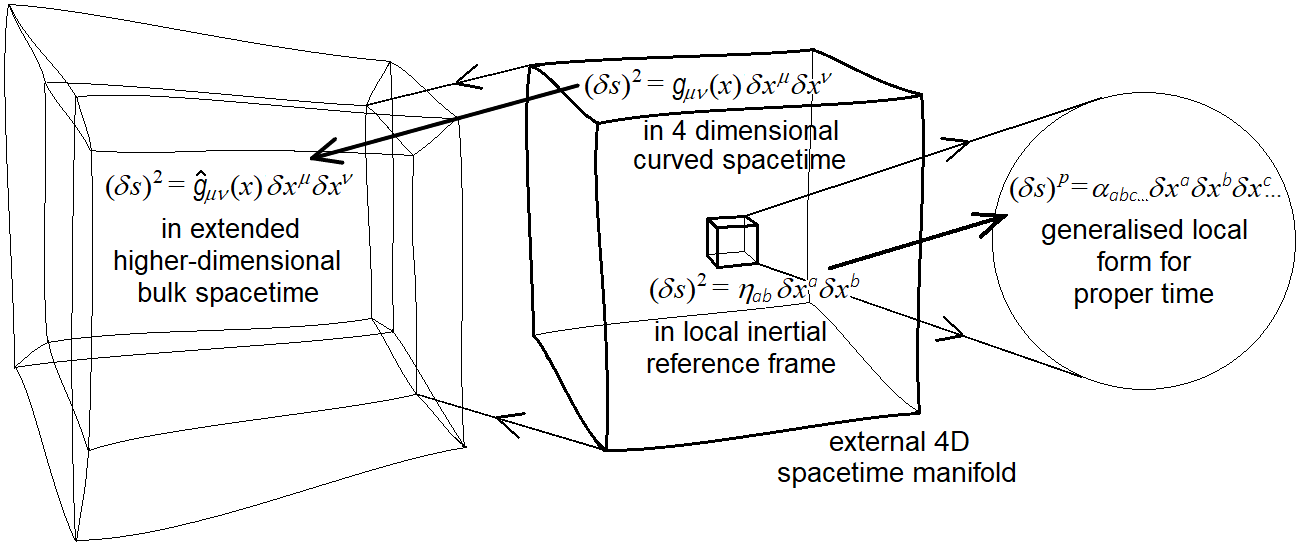}
\vspace{-35pt}    
\caption{\setb
 Contrasting global and local means of augmenting the 4-dimensional spacetime structure.
  In general relativity the quadratic spacetime form for an infinitesimal proper time interval $\delta s \in \rrr$ in a curved 4-dimensional spacetime can be expressed in general global coordinates $\{x^{\mu}\}$ via the symmetric metric tensor 
 $g_{\mu\nu}(x)$, with $\mu,\nu=0,1,2,3$, or in terms of a set of local inertial coordinate parameters $\{x^{a}\}$ and the 
  Lorentz metric $\eta_{ab} = \mbox{diag}(+1,-1,-1,-1)$, with $a,b=0,1,2,3$, in the form of equation~\ref{seta}. 
  For these global and local forms there are then contrasting possibilities for generalisation via additional components, as pictured to the left and right respectively, as a potential means of incorporating structures of matter.
   In models with extra spatial dimensions the global metric is augmented to  $\hat{g}_{\mu\nu}(x)$,
  now  with $\mu,\nu=0,\ldots,n-1$, in an $(n>4)$-dimensional extended `bulk' spacetime. An alternative generalisation can be made from the local quadratic form with local metric $\eta_{ab}$  to a $p^{\mathrm{th}}$-order homogeneous polynomial form for a proper time interval $\delta s$ in $(n>4)$ components as described for equation~\ref{salpha}.
   A key point for the new theory is that the additional components $\{\delta x^{a}; a>3 \}$
    are \textit{not} interpreted as coordinate intervals in an extended spacetime structure but rather provide a \textit{direct} basis for structures of matter.
   }
\label{contra}
\end{figure}

    Higher-dimensional spacetime models (see for example~\cite{Rubak,Liu}) typically augment the original global metric $g_{\mu\nu}(x)$  on introducing additional extended spatial dimensions. With the augmented metric denoted  $\hat{g}_{\mu\nu}(x)$ 
     in an $(n\!>\!4)$-dimensional  `bulk' spacetime  
      a proper time interval can still be expressed as a quadratic form, that is 
     $ (\delta s)^2 = \hat{g}_{\mu\nu}(x) \delta x^{\mu} \delta x^{\nu}$, as depicted on the left-hand side of figure~\ref{contra}.
      A \mbox{4-dimensional} spacetime base together with a matter content can then be extracted from the larger bulk spacetime structure, for example by `compactifying' the extra dimensions.

    Here we consider an alternative approach for which it is the local 4-dimensional quadratic form for proper time  with local metric $\eta_{ab}$
     that is generalised  to a $p^{\mathrm{th}}$-order  form for an infinitesimal  proper time interval $\delta s$ in $(n\!>\!4)$ components, that is:
\begin{eqnarray} 
  \mbox{from} \quad    (\delta s)^2  & = & \eta_{ab}\delta x^a \delta x^b  
     \quad \qquad \qquad\:\! \mbox{with} \quad a,b = 0,1,2,3  \label{seta}   \\               
 \mbox{to} \quad   (\delta s)^p  & = & \alpha_{abc\ldots}\delta x^a 
                            \delta x^b \delta x^c \ldots    
        \quad \mbox{with} \quad a,b,c,\ldots = 0,\ldots,n-1                     
                               \label{salpha}
\end{eqnarray}
   with homogeneous polynomial integer power $p\ge 2$, as depicted on the right-hand side of figure~\ref{contra}.  Each coefficient takes a value $\alpha_{abc\ldots} = -1,0$ or $+1$ in generalising the metric coefficients $\eta_{ab} = \mbox{diag}(+1,-1,-1,-1)$, with a sum in equation~\ref{salpha} over each index $a,b,c,\ldots$ for this $n$-component generalised form (consistent with the standard convention of summing over repeated upper and lower indices). While the proper time interval in equation~\ref{seta} is invariant under Lorentz transformations, with the Lorentz group $\soot$ acting on the four $\{\delta x^a\}$ components, the interval $\delta s$ in equation~\ref{salpha}
   in representing generalised proper time 
    is invariant under a larger symmetry group denoted $\hG$ (containing an $\soot$ subgroup),
     now acting upon a full set of $n$ $\{\delta x^a\}$ components collectively denoted
        $\bxn \in \rrr^n$. 
     In particular, since the whole purpose of augmenting from the 4-dimensional spacetime form is to incorporate a structure of \textit{matter}, rather than more \textit{space}, we can drop any restrictive assumption of a quadratic $p=2$ form for equation~\ref{salpha} as a local generalisation for the form of proper \textit{time}.

     As an augmentation from the local 4-dimensional spacetime form of equation~\ref{seta},
    the $n$-component form of  equation~\ref{salpha} can be decomposed as:
\begin{equation}
   \tag{3(a)}\label{sbreak}
  \underbrace{(\delta s)^p   =  \alpha_{abc\ldots}\delta x^a 
                            \delta x^b \delta x^c \ldots}_{\stackrel{
                                 \mbox{\small{$a,b,c,\ldots\, =\, 0,\ldots,n-1$}}}
                                          {\mbox{\raisebox{-9pt}{\small{generalised proper time}}}}}   =
                         \underbrace{( \eta_{ab}\delta x^a \delta x^b )}_
                         {\stackrel{ \mbox{\small{$a,b\, =\, 0,1,2,3$}}}
                         {\mbox{\raisebox{-9pt}{\small{spacetime basis}}}}}  
	  \!\!  \times   \underbrace{(\bxnf)^{p-2}  \,\;\,
    +  \,\;\,  (\bxn)^{p}}_
         {\stackrel{\mbox{\small{$\;\, \delta x^c;\, c\ge 4\qquad \;\;
            \delta x^a;\,  a\ge 0 \!\!$}}}
               {\mbox{\raisebox{-9pt}{\small{basis for matter}}}}}       \,
\end{equation}
        where $(\bxnf)^{p-2}$ and $ (\bxn)^{p}$ represent the appropriate polynomial expressions in the corresponding components (with $\bxnf \in \rrr^{n-4}$ a subset of $\bxn$) as consistent with equation~\ref{salpha}.
         The full symmetry $\hG$ of equation~\ref{salpha} and the left-hand side of equation~\ref{sbreak} is broken in \textit{extracting} four preferred components $\bdx_4 \equiv \{\delta x^a; a=0,1,2,3\} \in \rrr^4$ upon which the Lorentz $\soot \subset \hG$ symmetry acts, as the basis for the local external \mbox{4-dimensional} spacetime,   leaving a residual internal symmetry $G\subset \hG$. 
    This internal Lie group $G$ can be defined as the subgroup generated by the subset of the generators of $\hG$ that commute with the external Lorentz group generators in the corresponding Lie algebra. 
    The internal group $G$ is interpreted as a local gauge symmetry that will determine the possible local interactions of particle states. 
         In \mbox{4-dimensional} spacetime the full symmetry $\hG$ is hence broken down to the local product structure:      
\begin{equation}
  \color{white}  \label{extra}
\end{equation} 
\vspace{-41pt}
\begin{equation}
   \tag{3(b)} \label{gbreak}
    \mbox{Lorentz} \times G \subset \hG
\end{equation}

   The fragmented components of $\bxn$ in equation~\ref{sbreak}, transforming under 
   the surviving broken symmetry of equation~\ref{gbreak}, form the basis for the local structure of matter fields and physical particle states themselves,
   while the local gauge symmetry $G$ will be associated with the gauge bosons that mediate the local interactions between these particle states.    
    In all cases all the matter and gauge fields are identified through the symmetry breaking in
       equations~\ref{sbreak} and \ref{gbreak} with no further structure required to be posited.
       
          The full symmetry $\hG$ of equation~\ref{salpha} is not a symmetry of the physics, rather the symmetry breaking described for equations~\ref{sbreak} and \ref{gbreak} is
     \textit{required} in order to construct  \mbox{4-dimensional} spacetime itself and identify \textit{any} physics it contains. 
 Hence for example the surviving external and internal group product structure in equation~\ref{gbreak}, 
  as the symmetry for physical particle states, is consistent with the demands of the Coleman-Mandula theorem for any relativistic particle scattering theory
  that arises within this framework (see for example~\cite{TimeE} discussion of equation~83).

     Similarly as for theories based upon a higher-dimensional spacetime this local construction then provides a conceivable and credible means of deriving the elementary physical structures of matter via an augmented spacetime form. 
  Further, with reference to the contrasting perspectives described for figure~\ref{contra},  a global extension with extra dimensions of space, followed for example by a prescription of compactification to reduce the additional dimensions down to an imperceivable scale, provides a somewhat indirect and ambiguous means of identifying a local structure of matter in 4-dimensional spacetime. 
   By comparison 
   the local generalisation of proper time, with the basis for matter identified directly alongside the basis for local 4-dimensional spacetime through the composition of equations~\ref{sbreak} and \ref{gbreak}, 
   provides a very direct and unambiguous means of deriving local structures of matter.  
  While the conceptual basis of generalised proper time is hence well-motivated  in comparison with extra spatial dimensions the new approach also proves to be notably more successful in providing direct connections with both the Standard Model of particle physics and dark sector models in cosmology, as we describe in the following two sections.


\vspace{-7pt}
\section{Connections with the Standard Model}
\label{bas3}
\vspace{-1pt}

  The generalisation of the local form for a proper time interval leads directly to specific mathematical structures. 
    A unique sequence of augmentations from the \textit{quadratic} local 4-dimensional spacetime form of equation~\ref{seta}, consistent with the generalised form for proper time in  equation~\ref{salpha} with $p>2$, can be identified on 
    first expressing equation~\ref{seta} in the matrix form:
\vspace{-3pt}
\begin{equation}
  \label{sedet}
    (\delta s)^2
	 =  \eta_{ab}\delta x^a \delta x^b
	   = \det (\bdx_4) \quad \mbox{with} \quad
	    \bdx_4 =  \left(   
	   \begin{array}{cc} \delta x^0 + \delta x^3
	         & \delta x^1 - \delta x^2i \\
		   \delta x^1 + \delta x^2i  
		     & \delta x^0 - \delta x^3  \end{array} \right)  \in \htwc   
\end{equation}
  The Lorentz symmetry group $\soot$ of equation~\ref{seta} is correspondingly replaced by its double cover with elements $S\in \sltc$ acting upon the $2\times 2$ complex Hermitian matrices
     $\bdx_4 \in \htwc$  as
    $\bdx_4 \to S\bdx_4 S^{\dagger}$ in the standard way.

     Equation~\ref{sedet} can then be augmented directly to a \textit{cubic} form for proper time defined by the determinant of $3\times 3$ complex Hermitian matrices over
      nine real components denoted $\bdx_9$, that is:
\begin{equation}
  \label{scdet}
    (\delta s)^3
	   = \det (\bdx_9) \quad \mbox{with} \quad
	    \bdx_9   \in \hthc   
\end{equation}  
  with a full $\hG = \slthc$ symmetry. An external Lorentz $\sltc$ subgroup factor acts upon the four original components $\bdx_4$ extracted as the basis for the local \mbox{4-dimensional} spacetime as described for equation~\ref{sbreak}, with
   the full symmetry  broken to
   $\sltc \times \uo \subset \slthc$ and hence  leaving an internal $G=\uo$ symmetry for this case of equation~\ref{gbreak}. The five residual components, transforming under this broken symmetry, can be interpreted as the basis for a structure of matter as described in the previous section 
   (see also \cite{Gener} subsection~2.3).
   
     The complex algebra $\ccc$ as utilised in equation~\ref{scdet} can also be generalised to larger normed division algebras via the 4-component quaternions $\hhh$ to the 8-component octonions $\ooo$. Despite the non-associativity of the octonion algebra it is still possible to define a determinant for 
     $3\times 3$ octonion Hermitian matrices $\htho$ (\cite{Baez1} section~3.4), and hence a
      further \textit{cubic} form for a proper time interval now over 27 real components $\bdx_{27}$ can be defined as: 
\begin{equation}
  \label{sodet}
    (\delta s)^3
	   = \det (\bdx_{27}) \quad \mbox{with} \quad
	    \bdx_{27}   \in \htho   
\end{equation}        
   with a full $\hG = \sltho \equiv \esi$ symmetry~(\cite{Wang2} and references therein). Specifically, the
   
   \pagebreak
   \noindent
    corresponding $\sltho$ symmetry constructed is equivalent to the $\esig$ real form of the exceptional Lie group $\esi$.

    In turn this symmetry action is known to embed in an $\eseg$ symmetry that leaves invariant a quartic form $q$ defined over the 56-component space of the Freudenthal triple system $\fhtho$
    (such as described in~\cite{Baez1,Rios}). 
   This $4^{\mathrm{th}}$-order homogeneous polynomial form, as a further augmentation from equations~\ref{seta} and \ref{sedet} via equations~\ref{scdet} and \ref{sodet}, can then be interpreted as a \textit{quartic} form for generalised proper time over $n=56$ real components  $\bdx_{56}$:
\begin{equation}
  \label{sfq}
    (\delta s)^4
	   = q(\bdx_{56}) \quad \mbox{with} \quad
	    \bdx_{56}   \in \fhtho   
\end{equation}       
      with a $\hG=\ese$ exceptional Lie group symmetry.

        Proceeding as described for equation~\ref{sbreak}  on extracting the four components 
        $\bdx_{4}$ of the local external spacetime the residual components transform under the  broken symmetry found for equation~\ref{gbreak}. The internal symmetry $G=\uo$
        obtained for equation~\ref{scdet} can be augmented to
        $G=\suth_c\times\uo_Q \subset\sufo \subset \esi$
         for the $\hG = \esi$ symmetry level of equation~\ref{sodet}, with
     $c$ denoting colour and $Q$ electromagnetism on interpreting these internal gauge symmetries in the context of the Standard Model. 
      These assignments are motivated by the       
       symmetry breaking pattern obtained, which is used in turn as the basis for analysis of the $\ese$ level of equation~\ref{sfq} with the results described for table~\ref{esiese}.  
       \vspace{-1pt}
\def\rai{+0.2ex}
\begin{table}[htbp]
\centering
\begin{tabular}{|r|ccccc|c|}
 \hline
  \raisebox{-0.5ex}{${56}$} \!\!\!\!\!\!\!
   {\mbox{\raisebox{+0.0ex}{\LARGE{$\diagdown$}}}} \!\!\!\!\!\!
      \mbox{\raisebox{+0.7ex}{$\ese \! \supset\!\!$}} 
	    & \raisebox{\rai}{Lorentz} &
  \raisebox{\rai}{$\!\!\!\times\!\!\!$}
	    & \raisebox{\rai}{$\suth_c$}   &
  \raisebox{\rai}{$\!\!\!\times\!\!\!$} 
		& \raisebox{\rai}{$\uo_Q$} & \raisebox{\rai}{matter} \\
 \hline
    & & & & & &
   \vspace{-15pt} 
	\\	
       $ \!\!\!\!\!\!\!\!   {  \begin{array}{c}  \mbox{\raisebox{+0.3ex}{\small $\to$}} \\ 
                                        \mbox{\small}  \end{array} } \!\!\!\!\! $ 
	  $ \left(\!\!   {  \begin{array}{c}  \mbox{\normalsize$4$} \\ 
                                        \mbox{\normalsize$4$}  \end{array} } \!\!\right) \,   $  	
	 & 
	 $  \!\!    \begin{array}{c}  \mbox{\normalsize vector} \\ 
                                        \mbox{\normalsize  \underline{vector}}  \end{array}  \!\!$
	  & &  
	$  \!\!    \begin{array}{c}  \mbox{\normalsize$\bone$} \\ 
                                        \mbox{\normalsize  $\bone$}  \end{array}  \!\!$   
	   & &  
	   $  \!\!    \begin{array}{c}  \mbox{\normalsize $0$} \\ 
                                        \mbox{\normalsize  $0$}  \end{array}  \!\!$
	   & 
        $  \!\!    \begin{array}{c}  \mbox{\normalsize $\bdx_4$ `Higgs'} \\ 
                                        \mbox{\normalsize  `$\nu_L$'}  \end{array}  \!\!$     \\
   & & & & & &
    \vspace{-15pt} \\    	
      $ \!\!\!\!\!\!   { \scriptsize  \begin{array}{c}  \mbox{\raisebox{+0.3ex}{\small $\to$}} \\ 
                                        \mbox{\small}  \end{array} } \!\!\!\!\! $ 
	 {\small $ \left(\!\!   { \scriptsize \begin{array}{c}  \mbox{\normalsize$4$} \\ 
                                        \mbox{\normalsize$4$}  \end{array} } \!\!\right)$ } 	
	 & Dirac & & $\bone$  & & 1 & 
    {\small $ \left(\!\!   { \scriptsize \begin{array}{c}  \mbox{\normalsize$e_R$} \\ 
                                        \mbox{\normalsize$e_L$}  \end{array} } \!\!\right)$ }    \\
   & & & & & &
    \vspace{-15pt} \\   
      $ \!\!\!\!\!\!   { \scriptsize  \begin{array}{c}  \mbox{\raisebox{+0.3ex}{\small $\to$}} \\ 
                                        \mbox{\small}  \end{array} } \!\!\!\!\! $   	
    {\small $ \left(\!\!   { \scriptsize \begin{array}{c}  \mbox{\normalsize$6$} \\ 
                                        \mbox{\normalsize$6$}  \end{array} } \!\!\right)$ } 
	                 & \underline{scalar} & & $\bthr$ & & $\frac{2}{3}$ & 
  {\small $ \left(\!\!   {\scriptsize \begin{array}{c}  \mbox{\normalsize{`$u_R$'}} \\ 
                                        \mbox{\normalsize{`$u_L$'}}  \end{array} }  \!\!\right)$ }    \\ 
      & & & & & &
    \vspace{-15pt} \\  
      $ \!\!\!\!   { \scriptsize  \begin{array}{c}  \mbox{\raisebox{+0.3ex}{\small $\to$}} \\ 
                                        \mbox{\small}  \end{array} } \!\!\!\!\! $  
    {\small $ \left(\!\!   { \scriptsize \begin{array}{c}  \mbox{\normalsize$12$} \\ 
                                        \mbox{\normalsize$12$}  \end{array} } \!\!\right)   \!\!$ } 	
                                         & Dirac & & $\bthr$ & & $\frac{1}{3}$ & 
 {\small $\; \left(\!\!   {\scriptsize \begin{array}{c}  \mbox{\normalsize$d_R$} \\ 
                                        \mbox{\normalsize$d_L$}  \end{array} } \!\! \right) \!\!\: $ } 
                             \vspace{-15pt}               \\ 	
	  & & & & & &								 
					\\  
      & & & & & &
   \vspace{-22pt} 
	\\   
  $ \!\!\!\!\!\!   { \scriptsize  \begin{array}{c}  \mbox{\raisebox{+0.3ex}{\small $\to$}} \\ 
                                        \mbox{\small}  \end{array} } \!\!\!\!\! $ 
	 {\small $ \left(\!\!   { \scriptsize \begin{array}{c}  \mbox{\normalsize$1$} \\ 
                                        \mbox{\normalsize$1$}  \end{array} } \!\!\right)$ } 	
	 & scalar & & $\bone$  & & 0 &   
	  \\
   & & & & & &
    \vspace{-25pt} \\     
    $2 \quad\,$ & scalar & & $\bone$ & & 0 & 
	$\!\!\!\!$ 	\raisebox{0pt}[0pt][0pt]{\raisebox{+2.73ex}{$\Bigg\}$ 
	 {\tiny $\!\!\!\!    {\tiny \begin{array}{c}  \mbox{\normalsize Yukawa} \\ 
                                        \mbox{\normalsize coupling}  \end{array} } \!\!\!\!\!\!\!  $ } 	  
		 }}					
   	\vspace{-18pt}  \\  & & & & & &
	  \\
   \hline
  \end{tabular}
  \caption{\setb
   Symmetry breaking structure identified in the Standard Model sector of generalised proper time. The 
56 real components of $\bdx_{56} \in F(\htho)$ 
    for the $\ese$ quartic form  of equation~\ref{sfq} are partitioned under the broken symmetry on extracting a 
local Lorentz $\sltc$ subgroup acting upon the local external $\bdx_4$ 4-vector
subcomponents. The arrows indicate 
   the substructure fragmentation of the 27 real components at the $\esi$
    level of equation~\ref{sodet}. 
 The final column lists the matter state interpretation linking these structures with 
  the Standard Model (\protect\cite{TimeE} section~4).}
\label{esiese}
\end{table}       

\pagebreak

    Table~\ref{esiese} lists the fragmentation of both the $\bdx_{27}$ and   $\bdx_{56}$ components for equations~\ref{sodet} and \ref{sfq} respectively under the product of the external Lorentz $\sltc$ symmetry and the internal gauge symmetry $G=\suth_c\times\uo_Q$.
    The arrowed upper entries in the brackets in the left-hand column of 
         table~\ref{esiese} collectively  list the \mbox{27-component}
         substructure of the symmetry breaking for the intermediate $\hG=\esi$  level of equation~\ref{sodet}. These 27 components, corresponding to one $\htho$ subspace of $F(\htho)$, are considered to reside in the `right-handed sector' of the theory.  The $F(\htho)$ space at the $\ese$ level 
          contains a second $\htho$ subspace,  describing a `complex conjugate' representation of the 
          $\esi \subset \ese$ subgroup with respect to the first, with the corresponding 27 components residing in the `left-handed sector' and listed in the lower entries of the brackets in table~\ref{esiese}.
        The augmentation to the $\ese$ level of equation~\ref{sfq} then involves a doubling up of the $\esi$ 
   27-component case, with for example right-handed $R$ Weyl spinors accompanied by left-handed $L$ counterparts to form sets of Dirac spinors, together with two further scalar components as listed at the bottom of the table. 

 The central columns in table~\ref{esiese} result from the direct mathematical analysis of the symmetry breaking.  This structure is \textit{found} to be rich in Standard Model properties.
      Lorentz $\sltc$ Dirac spinor structures, $\suth_c$ singlets $\bone$ and triplets $\bthr$ and   $\uo_Q$ fractional charges, as listed by their relative magnitudes $\{1,\frac{2}{3},\frac{1}{3}\}$ in table~\ref{esiese}, motivate the associations with the Standard Model colour $\suth_c$ and
       electromagnetic $\uo_Q$ gauge symmetries together with the lepton 
        $(\nu,e)$ and quark $(u,d)$ states
      as indicated in the final column.

     Significant elements of an electroweak
     $\sutw_L \times \uo_Y$ symmetry structure (with $L$ left-handed and $Y$ hypercharge)  in relation to the internal $\suth_c \times \uo_Q$ symmetry in table~\ref{esiese} can be identified on temporarily omitting the external Lorentz component (\cite{TimeE} section~4). 
    These elements of electroweak symmetry breaking point to the external components $\bdx_4$, and the associated scalar
      $\det(\bdx_4)$, as being closely connected with the Standard Model Higgs, 
      with the contingent nature of this connection indicated by the quote marks on the
       `Higgs' assignment in the table.
    Further,  by comparison of terms in the expansion of the $\ese$ quartic form for proper time under the symmetry breaking with Standard Model Lagrangian mass terms, the scalars at the bottom of table~\ref{esiese} are provisionally associated with Yukawa couplings (see discussion of equation~\ref{extgpt} in section~\ref{bas5}).

         A left-right asymmetry arises from the asymmetric embedding of the external spacetime $\bdx_4$ components in the 56-component form for proper time in equation~\ref{sfq} and the
             extraction of the corresponding Lorentz $\sltc$ subgroup symmetry from the unification group $\hG = \ese$.
         As alluded to below table~\ref{esiese} 
            the $\mbox{\boldmath $56$}$ representation of $\ese$ associated with equation~\ref{sfq} incorporates the   $\mbox{\boldmath $27$}$ and {$\mbox{\boldmath $\overline{27}$}$} 
       representations of an $\esi$ subgroup structure, as expressed by two copies of $\htho$
        inside $\fhtho$ and corresponding to `right-handed' and `left-handed' sectors.
                     It is the necessary extraction of the external spacetime components $\bdx_4$ from one, and only one, of these two  complementary sectors that breaks the left-right symmetry
          (see also \cite{Gener} subsection~3.1).

         Overall, these structures identified in this very direct symmetry breaking for generalised proper time manifest a significant resemblance to  structures of the Standard Model.
         Some of the above observations might be anticipated since $\esi$ and $\ese$ form part of a well-known sequence of groups of interest in 
   Grand Unified Theories  (GUTs) that focus purely on the internal symmetry properties (\cite{PDG22} section~93.2.4).
   However, unlike the case for GUTs, here for table~\ref{esiese} the external Lorentz symmetry is extracted first from these unifying symmetries $\hG= \esi$ and $\hG=\ese$ of generalised proper time, as described for equations~\ref{sbreak} and \ref{gbreak}.  
   This leaves insufficient room to identify a fully realistic Standard Model internal gauge symmetry and corresponding one generation multiplet structure for these unification groups.

          Moreover,  in addition to  the requirement of incorporating the full Standard Model  gauge symmetry including a complete electroweak structure \textit{alongside} the external Lorentz symmetry,
                  the discrepancies in the spinor structure as underlined in table~\ref{esiese}
          (and prompting the quote marks on the `$\nu_L$'-neutrino and `$u_{R,L}$'-quark states) 
           and the need to describe a full three generations of lepton and quark states, firmly imply that a yet further augmentation from the $\ese$ level in table~\ref{esiese} should be sought.

    Given these empirical requirements  in augmenting from the 56-dimensional case for $\ese$ in table~\ref{esiese} the largest exceptional Lie group $\ee$, for which the smallest non-trivial representation is 248-dimensional, might naturally be suggested as a unique and potentially ultimate symmetry for a generalised form of proper time. There is indeed a known mathematical augmentation to a realisation of $\ee$ as a symmetry of the `extended Freudenthal triple system' (\cite{Rios} section~3.5), however that is a space with only 57 components.
      Here the question is whether a $\hG=\ee$ symmetry acting upon a homogeneous polynomial form for proper time consistent with equation~\ref{salpha}, as a final extension to the series of equations~\ref{sedet}--\ref{sfq}, might be identified. Consideration of these and other relevant mathematical structures (such as the existence of an $\ee$ octic invariant~\cite{CedP}) motivates the  proposal of a potential 
       $8^{\mathrm{th}}$-order form for generalised proper time:
\begin{equation}
  \label{soctic}
    (\delta s)^8
	   = Q(\bdx_{248}) \quad \mbox{with} \quad
	    \bdx_{248}   \in \mathcal{T}   
\end{equation} 
   where $Q$ is an octic form defined on the 248 real components of $\bdx_{248}$ as an element in a 248-dimensional space denoted $\mathcal{T}$, and with a full
    \mbox{$\hG = \eeg$} symmetry~\cite{TimeE}.  The aim then would be to determine the resulting augmentation from table~\ref{esiese} under a symmetry breaking for equations~\ref{sbreak} and \ref{gbreak} to be sought in the form:
\begin{equation}
  \label{ebreak}
       \mbox{Lorentz} \times \suth_c \times \sutw_L \times \uo_Y \subset \ee 
\end{equation}

  The mathematical structure and symmetry breaking of equation~\ref{soctic}  should explain why the internal gauge symmetry identified directly has this form with the three Standard Model component parts
 in equation~\ref{ebreak}  (rather than yielding a single internal `GUT' group that requires further breaking). That is, the theory should account for an internal $G=\suth_c \times \sutw_L \times \uo_Y$ as identified in the product structure of equation~\ref{gbreak} and incorporate in turn the electroweak breaking
   $\sutw_L \times \uo_Y \to \uo_Q$, with respect to the $\bdx_4$ `Higgs' components, hence subsuming the internal symmetry identified in table~\ref{esiese} for the $\hG=\esi$ and $\hG=\ese$ substructure levels.

      However, it is well known from the standard analysis of simple Lie algebras that the branching of the 248-dimensional representation of $\ee$ under the  broken symmetry of equation~\ref{ebreak} \textit{cannot} accommodate three generations of Standard Model states~\cite{Lisi,DiGa}.
      In~\cite{Lisi} an $\ff \times \gt \subset \ee$ substructure is employed with an
       $\soe \subset \ff$ subgroup containing both the external Lorentz $\sutw \times \sutw$ (in terms of the complexified structure) and internal weak $\sutw_L$ symmetries. In attempting to identify three  generations of Standard Model states in the branching pattern of the $\mbox{\boldmath $248$}$ representation under equation~\ref{ebreak} only the first is essentially complete, while the representations under these three $\sutw$s are mutually permuted by `$\soe$ triality' maps giving an incorrect structure for the second and third generations~(\cite{Lisi} for example section~2.4.2). This \textit{cannot} be fixed by analysis within standard Lie algebra theory since no real form of $\ee$ contains a sufficient number of non-compact generators to describe three generations of Standard Model spinor states (\cite{DiGa} section~3). Nevertheless such analyses \textit{do} indicate that a symmetry breaking pattern for the $\mbox{\boldmath $248$}$ representation of $\ee$ \textit{can} indeed be in the proximity of describing the Standard Model, to within the above significant $\soe$ triality issue.

     In the present theory the primary source of symmetries is through transformations upon homogeneous polynomial forms leaving the proper time interval $\delta s$ in equation~\ref{salpha} invariant, rather than directly via the standard classification and representation structure of simple Lie groups.
       In particular, for the cubic and quartic forms of equations~\ref{sodet} and \ref{sfq} the octonion algebra is key to the construction of both the representation space and the symmetry actions, and in turn octonions are also anticipated to be central to the explicit construction for the $\ee$ level of equation~\ref{soctic}. Hence a non-standard representation space  $\mathcal{T}$ can be obtained, with the non-associativity of the octonions also implying symmetry compositions that may not in general directly form a group structure (for which
        composition associativity is a key axiom).

         Hence with the construction heavily involving the octonion algebra the symmetry breaking pattern obtained from equations~\ref{soctic} and \ref{ebreak}  can deviate to some degree 
        from the expectations of standard  Lie group theory analysis, with explicit calculation required for these octonionic structures as has been the case for the $\esi$ and $\ese$ levels in constructing table~\ref{esiese}.
            In particular,  the property of `octonion triality', and its relation to $\soe$ and spinor representations
  (see for example~\cite{Baez1}), might in principle be found to compensate for the above-mentioned discrepancy involving an $\soe$ triality and lead to the correct identification of a full three generations of Standard Model states from the 248 real components of $\bdx_{248} \in \mathcal{T}$ in equation~\ref{soctic} under the symmetry breaking of equation~\ref{ebreak}.

        While grounded in the underlying conceptual basis for this theory in generalised proper time as described for figure~\ref{contra} and equation~\ref{salpha},  the existence of such an octonion-based $\ee$ structure and symmetry breaking pattern that accommodates the full set of Standard Model states constitutes a well-defined mathematical prediction for this approach~\cite{TimeE,Octo}. 
      This ambition bears some relation to ongoing developments concerning the 
  connections between the octonions, $\ee$ and the Standard Model that are already apparent from a purely algebraic point of view (see for example~\cite{Wils,Mano} and references therein).

    If such an $\ee$ symmetry and breaking pattern might be realised for equations~\ref{soctic} and \ref{ebreak} to reproduce the properties of a full three generations of Standard Model states this should also incorporate the basis for the significant level of mixing between the generations, in both the quark and neutrino sectors, as empirically observed. 
    The corresponding mismatch between weak interaction states and mass states, and the need for mass matrices spanning the three generations, implies that it may well \textit{not} be possible to mathematically extract a substructure describing a single pristine generation of quarks and leptons in isolation from a subset of the  $\bdx_{248} \in \mathcal{T}$ components in equation~\ref{soctic}.
      The apparent `defects' at the approximately one-generation level of  table~\ref{esiese} for the
      $\hG=\ese$ stage then reflect the somewhat `jagged edges' that might be expected to result from any  attempt to detach the substructure of a single generation. 
    
    \pagebreak
    
    In the context of the present theory the mixing relations between the three generations are likely to involve the above-mentioned octonion triality property in an intrinsic manner that would not be manifest in any potential one-generation substructure.
     Hence the identification of a  single `complete'  generation, with no mixing terms, is \textit{not} anticipated at any intermediate stage in developing the theory, and indeed such a structure \textit{does not} correspond to anything we observe in nature. On the other hand, models that set out with the aim of identifying the ideal of a single flawless generation (see for example~\cite{Kras}) are not necessarily on the right course, since such a construction neither exists in nature nor sits without significant modification within the desired full picture with three heavily mutually interrelated generations. 
     
     A further recent observation regarding the construction of equation~\ref{soctic} required to accommodate  three related generations concerns the Koide mass formula, according to which the empirically determined masses of the three charged leptons are found to be related as 
      $(m_e + m_{\mu} + m_{\tau})/(\sqrt{m_e} + \sqrt{m_{\mu}} +
        \sqrt{m_{\tau}})^2 \simeq \frac{2}{3}$
        to within around one part in $10^5$. 
        For the present theory with mass terms proposed to derive from terms of the octic form in equation~\ref{soctic} under the symmetry breaking, the Yukawa couplings for the charged leptons will typically be composed of a product of four factors that are mutually related in deriving from the same original set of 248 components. In turn models constructed to account for the  Koide mass formula (see for example~\cite{EMa,Sumi}), together with identities involving sums of squares in the octonion algebra, may provide further input guiding the mathematical construction sought for equation~\ref{soctic} as consistent with a full three generations of Standard Model states.

    Although  a specific mathematical form for equation~\ref{soctic} with an $\ee$-type symmetry and its breaking pattern may need to be realised before detailed physical predictions might be determined, in extrapolating from the
     $\esi$ and $\ese$ levels of table~\ref{esiese} preliminary expectations for new physics can already be deduced.
      In particular 
     with the `Higgs' $\bdx_4$  and left-handed neutrino `$\nu_L$' states provisionally associated with complementary components in the two $\htho$ subspaces as embedded
      within $\fhtho$ at the $\ese$ level, as listed
      in the top rows of table~\ref{esiese}, 
  it is suggested that in any augmentation to a three-generation structure with $3\times \nu_L$ states the Higgs will be effectively embedded in a right-handed neutrino $\nu_R$ sector.

   With spinor components also needing to be identified for the neutrino states under this augmentation, it is proposed that the physical scalar Higgs state, while associated with the 
   external \mbox{4-vector} $\bdx_4$ components, may be a non-elementary composite state as effectively composed from some of the right-handed neutrino spinor degrees of freedom, leaving no more than $2\times \nu_R$ physical states, consistent with a
     number of models~\cite{Gener,Ibar,KrHi}.
     As for those related models this can lead to predictions that could be testable in the laboratory.
       For example a 
      composite Higgs, or other form of non-standard or extended Higgs sector, is a prime candidate for new physics that may account for
       the $7\sigma$ discrepancy between the recent $W^{\pm}$ gauge boson mass
       measurement and Standard Model electroweak theory expectation~\cite{CDFW}.

    While $\ese$ and $\ee$ are of limited interest as GUT groups due to their lack of complex representations (\cite{PDG22} section~93.2.4), for the present theory as described above for the $\ese$ case of table~\ref{esiese} a left-right asymmetry arises from the extraction of  the Lorentz subgroup symmetry acting upon the external spacetime $\bdx_4$ subcomponents as embedded asymmetrically in the `right-handed sector' of the $\bdx_{56}$ components.
       Since the `Higgs' $\equiv \bdx_4$ components are associated with the generation of mass the asymmetric embedding in the neutrino sector could also account for the apparent significant empirical mass difference between $\nu_L$ and $\nu_R$ states, as well as provide the basis for the left-right asymmetry of weak interactions more generally.

        The sequence of forms for generalised proper time in equations~\ref{sedet}--\ref{soctic} through to the
     proposed $\hG=\ee$ symmetry with a central role for the octonions, hence involving both the largest exceptional Lie group and the largest normed division algebra, describes a unique augmentation beyond the local 4-dimensional spacetime form of equation~\ref{seta} as consistent with the generalised form for proper time in equation~\ref{salpha} for $p>2$.  
      These structures have been found to make highly non-trivial connections with the Standard Model and new physics beyond, as reviewed above.

      Predictions for new physics based upon an explicit form for equation~\ref{soctic}, in the Higgs and neutrino sectors and elsewhere, would be robust in the sense that they are established on top of the reconstruction of the full Standard Model itself while having a firm underlying conceptual basis in generalised proper time as described in the previous section. 
      However, equation~\ref{salpha} itself as the basis for this theory   is of a more general character, permitting further mathematically  possible `branches' for the augmentation of local proper time intervals and implying the  possibility of further forms of matter that can be identified  \textit{in parallel with} the Standard Model branch reviewed in this section. Through such structures  this theory  has a significant potential to also account for the dark sector as we describe in the following section.


\section{Connections with Dark Sector Models}
\label{bas4}
      
       In achieving the above connections with the Standard Model we described how
        rewriting the original 4-dimensional quadratic spacetime form of equation~\ref{seta}
        in the $2\times 2$
       matrix form of equation~\ref{sedet} allowed a natural augmentation of the form for proper time to a cubic $3\times 3$
       matrix expression with a $\hG=\slthc$ symmetry in equation~\ref{scdet}, leading  
        on directly to higher-dimensional forms with exceptional Lie group symmetries as described for
       equations~\ref{sodet}--\ref{soctic}. However, the $2\times 2 \to 3\times 3$ matrix structure extension could instead be further augmented to the 
        $p^{\mathrm{th}}$-order   
 determinant of $p\times p$ complex Hermitian matrices for any integer $p>3$:
\begin{equation}
  \label{spdet}
    (\delta s)^p
	   = \det (\bdx_{p^2}) \quad \mbox{with} \quad
	    \bdx_{p^2}   \in \hpc   
\end{equation}  
  with a full $\hG = \slpc$ symmetry acting upon the $n=p^2$ real components of $\bdx_{p^2}$, as consistent with the generalised form for proper time in equation~\ref{salpha} and as can be taken to the $p\to \infty$ limit.
  
  The extraction of the original
   external 4-dimensional spacetime components $\bdx_4$ of equation~\ref{sedet}, in common with the Standard Model branch 
   described for table~\ref{esiese}, 
   results in a broken symmetry for the branch of equation~\ref{spdet}, proceeding as described for  equations~\ref{sbreak} and \ref{gbreak}, of the form:
\begin{equation}
  \label{spbrk}
        \mbox{Lorentz}  \times \slptc  \times  \uod \subset \slpc
\end{equation}  
   with $k=p-2$. Since the internal symmetry $G = \slptc  \times  \uod$  is \textit{independent} of that of the Standard Model branch, described for the $\ese$ level of table~\ref{esiese} and as proposed  for the ultimate
     $\ee$  level of equations~\ref{soctic} and \ref{ebreak}, there are \textit{no local gauge interactions} in common with the visible matter sector. The structures of matter deriving from equation~\ref{spdet}  will then give rise to a `dark sector',  as denoted by the subscripts $D$ in equation~\ref{spbrk}.
        The corresponding symmetry breaking pattern for the fragmented components of the form of generalised proper time in equation~\ref{spdet} 
       is listed in table~\ref{ppdet}.
    

\def\rai{+0.2ex}
\begin{table}[htbp]
\centering
\begin{tabular}{|l|ccccc|c|}
 \hline
  \raisebox{-0.5ex}{${p^2\;}$} \!\!\!\!\!\!\!
   {\mbox{\raisebox{+0.0ex}{\LARGE{$\diagdown$}}}} \!\!\!\!\!\!
      \mbox{\raisebox{+0.7ex}{$\slpc \!\, \supset\!\!\,$}} 
	    & \raisebox{\rai}{Lorentz} &
  \raisebox{\rai}{$\!\!\times\!\!\!\!$}
	    & \raisebox{\rai}{$\slptc$}   & 
	 \raisebox{\rai}{$\!\!\times\!\!\!\!$}
	    & \raisebox{\rai}{$\uod$}   &        
	    \raisebox{\rai}{matter} \\
 \hline
     $\;\, 4 \qquad$ & 4-vector & & invariant & & 0 &  $\bdx_4$    external 				
	\\
  $\, k^2\;\, (k=p-2)$ & scalar  & &
   $\bdx_{k^2} \to  S_k\:\! \bdx_{k^2}\:\! S^{\dagger}_k$ 
     & & 0 &  $\bdx_{k^2}$ dark  $\bphi$
	\\
  $4k$ & Weyl  & & 
       $\bdx_{4k} \to S_k\:\! \bdx_{4k}$
     & & 1 &  $\bdx_{4k}$ dark    $\bpsi$	
	  \\
   \hline
  \end{tabular}
  \caption{\setb   
  Symmetry breaking structure for a proposed dark sector branch of generalised proper time.
  The full $\hG = \slpc$ symmetry of  equation~\ref{spdet} is broken in the projection over the local external  4-dimensional spacetime subcomponents $\bdx_4$ as described for equation~\ref{spbrk}.
   The resulting basis for
  matter states includes a set of $k^2$ scalars $\bphi$ (with $k=p-2$), transforming as the matrix $\bdx_{k^2} \in \hkc$ under determinant-preserving internal 
    $S_k \in\slptc$ actions and neutral under $\uod$,  together with a set of \mbox{$k$ 2-component} complex  Weyl spinors $\bpsi$,  transforming under the standard representation of $\slptc$ acting on
        the $k\times 2$ complex matrix $\bdx_{4k}$  and  charged under the internal dark $\uod$.
         As a further feature the non-Abelian 
    internal gauge symmetry $\slptc$ is \textit{non-compact} for $k=p-2\ge 2$, 
          supplying the link with dark energy models (as initially investigated in~\protect\cite{DEner}). }
\label{ppdet}
\end{table}

   For a non-Abelian gauge theory there are  a set of gauge field components $A^{\alpha}_{\pha{o}\mu}(x)$ with
    \mbox{$\alpha = 1, \ldots, n_G$}, where $n_G$ is the dimension of the Lie algebra. The corresponding gauge field strength $F^{\alpha}_{\pha{o}\mu\nu}(x)$,
     Lagrangian $\lag_{\Yml}$ and energy-momentum $T^{\mu\nu}_{\!\Yml}(x)$ are respectively 
    (the subscript `YM' denotes Yang-Mills, $\alpha,\beta,\gamma$ are indices in the Lie algebra and
      $\cabg$ the associated structure constants, while $\mu,\nu,\rho,\sigma$ are 4-dimensional spacetime coordinate indices):
 \begin{eqnarray}     
       F^{\alpha}_{\pha{o}\mu\nu}  & = &
        \pal_{\mu}A^{\alpha}_{\pha{o}\nu} - \pal_{\nu}A^{\alpha}_{\pha{o}\mu}
         + \cabg A^{\beta}_{\pha{o}\mu}A^{\gamma}_{\pha{o}\nu}
         \raisebox{-10pt}{{ }}
          \label{fym} \\
       \lag_{\Yml} & = & \!\! -\frac{1}{4}  
         F^{\alpha}_{\pha{o}\mu\nu}F_{\alpha}^{\pha{o}\mu\nu} 
            \qquad\!    =  \;  -\frac{1}{4}  K_{\alpha\beta}
         F^{\alpha}_{\pha{o}\mu\nu}F^{\beta\mu\nu}  
          \raisebox{-10pt}{{ }}
          \label{lagym}  \\
       T^{\mu\nu}_{\!\Yml} & = &  + F^{\alpha\mu}_{\pha{ow}\rho}
                                               F_{\alpha}^{\pha{o}\rho\nu} +
        \frac{1}{4} g^{\mu\nu}  F^{\alpha}_{\pha{o}\rho\sigma} 
                                             F_{\alpha}^{\pha{o}\rho\sigma}  \label{tym}
\end{eqnarray}        
      In particular equation~\ref{lagym} is considered the appropriate gauge and Lorentz invariant generalisation from the Maxwell Lagrangian $\lag_{\Mxl}  =  -\frac{1}{4} F_{\mu\nu}F^{\mu\nu}$, representing the kinetic energy term for the electromagnetic field, to the non-Abelian case of Yang-Mills gauge theory.

    By definition for a non-compact gauge symmetry, such as $\slptc$ in equation~\ref{spbrk} and table~\ref{ppdet},  the
     Killing metric $K_{\alpha\beta}$ of the Lie algebra in equation~\ref{lagym} is \textit{indefinite}. That is,  Lie algebra bases can be found with each element $K_{\alpha\beta} = \pm \delta_{\alpha\beta}$ but there is no basis in which as a whole $K_{\alpha\beta} = + \delta_{\alpha\beta}$.
   This  
  implies  an inevitable source of negative kinetic energy without a lower bound and a seeming instability of the vacuum in the corresponding quantum field theory 
  (see for example~\cite{WittQ} discussion of equation~1.1). 
  For this reason non-compact gauge symmetry groups are generally avoided in model building and this branch of generalised proper time of equation~\ref{spdet} and table~\ref{ppdet} would seem to be problematic or simply non-physical.

   However, a source of negative kinetic energy is a central ingredient in models
    for `phantom dark energy'~\cite{Caldw,Kujat,Ludw}.
      A `wrong-sign' kinetic term
     $\lag_{\phi} = - \frac{1}{2}\partial_{\mu}\phi\:\! \partial^{\mu}\phi$, together with a suitable potential term $V(\phi)$, is proposed for a postulated
      scalar field $\phi(x)$ in such models. In this paper we are then considering whether the negative kinetic energy contribution implicit for a non-compact gauge group might also be related to dark energy and the observed large-scale accelerating expansion of the universe.

      For   Friedmann-Lema{\^i}tre-Robertson-Walker (FLRW)  cosmological models gravity on large scales is assumed to be described by the metric  tensor  $g_{\mu\nu}(x)$ of general relativity and the Einstein field equation:
    \begin{equation}
   \label{Einel}
       G^{\mu\nu} + \Lambda g^{\mu\nu} = -\kappa T^{\mu\nu}
 \end{equation}    
     This equation relates the Einstein tensor  $G^{\mu\nu}(x)$, a non-linear function in the spacetime 
        derivatives of the metric field  $g_{\mu\nu}(x)$,
         to the energy-momentum tensor $T^{\mu\nu}(x)$, describing the matter content of the universe,  
    with $\kappa$ a positive normalisation constant. 
     In the present theory, while the structures of matter deriving from different branches of generalised proper time, such as described for tables~\ref{esiese} and \ref{ppdet}, are independent in terms of local gauge interactions, in sharing a common 4-dimensional external spacetime base these sectors will mutually interact through the classical gravitation of equation~\ref{Einel}.

    In FLRW cosmological models the cosmological principle concerning large-scale homogeneity and isotropy is also assumed.  
    The most general form of energy-momentum consistent with the cosmological principle can be written as:
\begin{equation}
  \label{tperf}
     T^{\mu\nu} = (\rho + p)u^{\mu}u^{\nu} - pg^{\mu\nu}
\end{equation}  
     Here the 4-velocity vector $u^{\mu}(x)$ represents the idealised average uniform flow of galaxies in the large-scale universe, as usually expressed in a simple form with components $(1,0,0,0)$ in a comoving
      cosmological reference frame, with the first coordinate being the cosmic time $t$.   
       The parameter $\rho$ ($\equiv T^{0}_{\pha{0}0}$) represents the energy density
       while
      \mbox{$p$ ($\equiv -T^{1}_{\pha{1}1} = -T^{2}_{\pha{2}2} = -T^{3}_{\pha{3}3}$)}  is interpreted as pressure, both of which are uniform in space at any given cosmic time $t$  (in a more general context, equation~\ref{tperf} describes a `perfect fluid').
      In addition to the energy density $\rho$ the 
      pressure $p$ of a fluid or substance also contributes to its gravitational impact in general relativity.  
    
    A positive cosmological constant $\Lambda$ on the left-hand side of the Einstein field equation~\ref{Einel}, in itself driving a cosmic acceleration,  is equivalent to a positive energy density $\rho_{\Lam}$ and negative  pressure 
    $p_{\Lam}$ on the right-hand side, with $p_{\Lam} = w_{\Lam} \rho_{\Lam}$ and an equation of state parameter $w_{\Lam} =-1$. Such a special case for dark energy is proposed in `the standard  cosmological model', or `$\Lambda$CDM', which incorporates also a significant cold dark matter (CDM) contribution as consistent with observations (\cite{PDG22} sections~22.1 and 25.1). 
      More generally, a source of dark energy with an equation of state with parameter $w<-\frac{1}{3}$ is required to describe a cosmic acceleration. 
    
    Here we consider the contribution to the energy-momentum $T_{\mu\nu}(x)$ on the right-hand side of equation~\ref{Einel} due to the non-compact non-Abelian gauge structure of 
    
    \pagebreak
    \noindent
    table~\ref{ppdet}.
     $\Apls(x)$ will denote the gauge field component with positive kinetic energy, 
      as associated with all diagonal Killing metric elements with 
      $K_{\alpha\alpha} = +1$ (no sum over $\alpha$), while $\Amin(x)$ will denote the remaining negative kinetic energy
      contribution, as associated with generators of the Lie algebra with corresponding 
        $K_{\alpha\alpha} = -1$ components in
       equation~\ref{lagym}.
      (More generally, `$+$' or `$-$' subscripts on fields in this paper will always denote positive or
       negative kinetic energy respectively).

       In fact the $\Apls$ states are associated with both positive energy density $\rAp>0$ and positive pressure $\pAp >0$. On reversing the sign of the implicit $K_{\alpha\alpha}$ factors in equation~\ref{tym} from $+1$ to $-1$ it can be seen that the  $\Amin$ states will be associated with a complementary negative energy density $\rAm <0$ as well as 
        negative pressure $\pAm < 0$.  We note that a contribution of \textit{negative pressure} is an essential component for any candidate for dark energy consistent with observations.
       Further, by the symmetry between the  $K_{\alpha\alpha} = +1$ and $K_{\alpha\alpha} = -1$ components in this construction we expect the corresponding positive and negative energy contributions to be accompanied by equivalent but contrary pressure contributions of the form:
 \begin{equation}
  \begin{array}{rcl}
          \pAp  = \wAp \rAp >0   & \!\!\!\! &  \;\; \mbox{for} \; \Apls(x) \; \mbox{fields}    
                            \\  
         \pAm = \wAm \rAm \, <0   & \!\!\!\! &  \;\; \mbox{for} \,\, \Amin(x) \:\, \mbox{fields}    
                            \\
          \mbox{with} \;\;  \wAp  &\!\!\!\!\!  =  \!\!\!\! &    \wAm > 0   
     \end{array}                 
         \label{pwppm}
\end{equation}

     A determination is required of the  large-scale  average effective
       $\pvac$ and $\rvac$ `vacuum'
      contribution terms for equation~\ref{tperf}  and their impact in the context of an FLRW cosmological model for this branch of equation~\ref{spdet} and table~\ref{ppdet}.
       The corresponding effective cosmological equation of state
        \mbox{$\pvac = \wvac \rvac$}, as potentially underlying the cosmic acceleration
      and in relation to   $\Lambda$CDM, 
      might then be fully assessed. 
       With the effective `vacuum' corresponding to the lowest \textit{net} energy state 
   any determination of $\pvac$ and $\rvac$  should incorporate  a combination of both the positive energy $\Apls(x)$ and
   negative energy $\Amin(x)$  field contributions from equations~\ref{pwppm}.

      For this non-Abelian gauge group component terms of cubic and quartic order in the gauge fields in 
      equations~\ref{lagym} and \ref{tym}, arising from the final term in equation~\ref{fym} for the field strength,   
       imply a significant degree of \textit{self-interaction} for the gauge fields from an apparent `kinetic energy' term. 
  Hence in particular the
     positive energy gauge bosons $\Apls$ together with the negative energy   gauge bosons $\Amin$ can \textit{mutually} interact. This implies a \textit{limitless} energy-momentum conserving generation of 
       $\Apls$ and $\Amin$ states out of the vacuum, via processes such as depicted in figure~\ref{AAonly}(a), from within this proposed dark energy sector alone.      
      
       However,  the spontaneous creation of these states throughout spacetime, via such processes  would result in a \textit{sea} of  $\Apls$ and $\Amin$ gauge   bosons and a degree of \textit{mutual annihilation}  would inevitably ensue, removing some of the positive and negative energy contributions. With the rate at which positive and negative energy is taken out of the system through this annihilation increasing as the magnitude of the respective density contributions 
       $\rAp$ and $\rAm$ increases, and assuming a constant rate of creation, a stable equilibrium might be rapidly attained through the cubic and quartic self-couplings,  
         as represented in figure~\ref{AAonly}(b).
      
\pagebreak          
       
\begin{figure}[htbp]    
\centering
\leavevmode
\includegraphics[width=14.4cm]{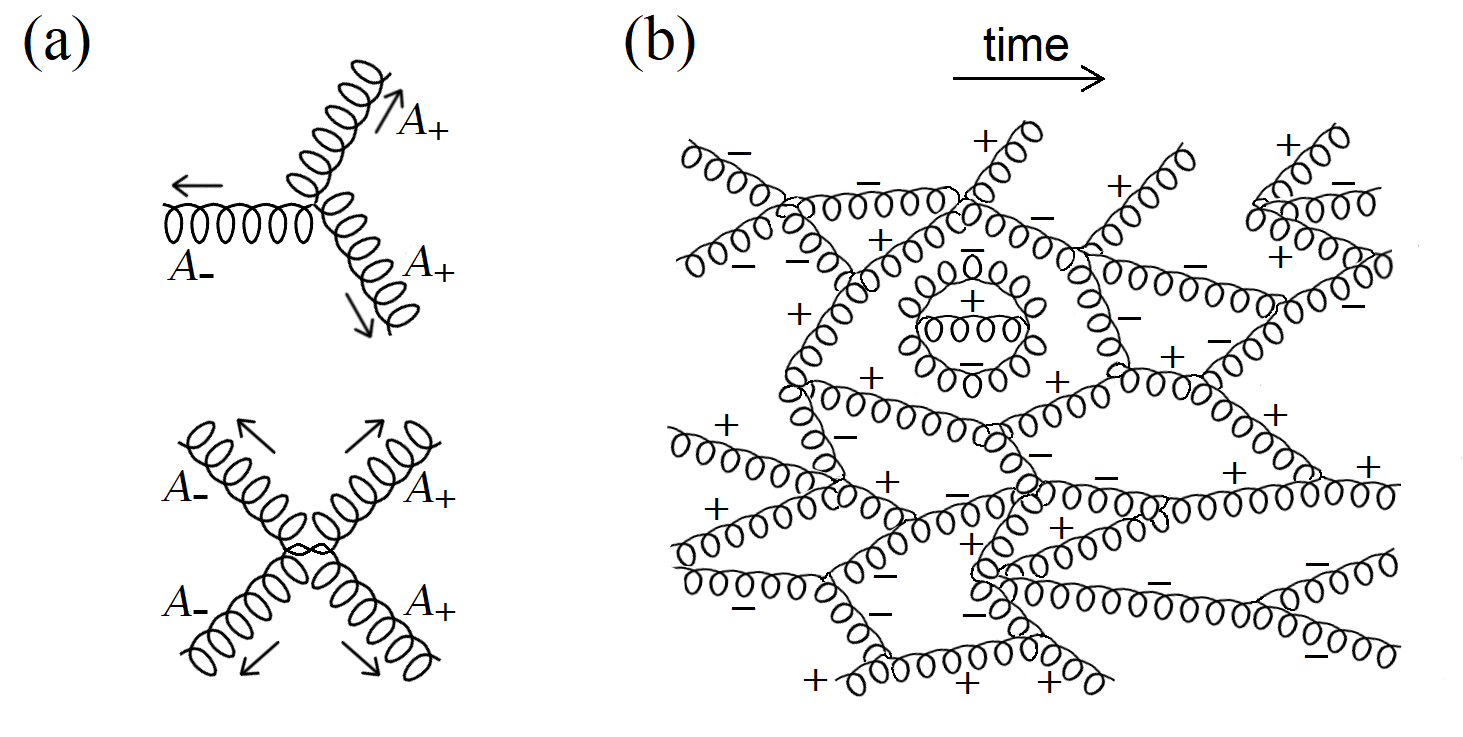}
\caption{\setb
   For a non-compact non-Abelian gauge sector the vacuum rapidly fills with positive and negative gauge boson states.
    (a) Production of negative energy $\Amin$ and positive energy $\Apls$ gauge bosons of a non-compact gauge symmetry out of the vacuum, as mediated directly through the self-interactions for the non-Abelian gauge symmetry (as kinematically permitted). As distinct from more typical Feynman diagrams here the arrows indicate the forward direction of time and the creation of both positive and (literally) negative energy states.
     (b) The dark vacuum is filled by a sea of $\Apls$ and $\Amin$ gauge bosons (the coiled 
   lines labelled `$+$' and `$-$' for the positive and negative energy states) through these processes, now depicted with time flowing from left to right for all states. A stable situation may arise in which the creation of these  states is balanced by their mutual annihilation, with net zero pressure and energy density as described for equations~\ref{prnona}.
     }
\label{AAonly}
\end{figure}

     As noted above in constructing equations~\ref{pwppm}, from equations~\ref{fym}--\ref{tym}
         there is a \textit{symmetry} between
         the $\Apls$  $K_{\alpha\alpha}=+1$ and $\Amin$  $K_{\alpha\alpha}=-1$
          contributions    that suggests the
             corresponding positive and negative  contributions, to both pressure and energy density, 
        would be equivalent and complementary,  individually satisfying the same 
          \textit{effective} equation of state for which  
    $\wAp = \wAm$ can be assumed.
          An equilibrium ground state for
           figure~\ref{AAonly}(b) could then be achieved with  the symmetric properties 
           $\pAp=-\pAm>0$ and  $\rAp=-\rAm>0$. Hence
      the net impact would be expected to \textit{cancel},  with the \textit{total} pressure and energy density for the vacuum state of this non-compact gauge sector vanishing:
\begin{eqnarray}
       \pvac &= & \;\; \pAp  \quad  + \quad \pAm \;\; = \; 0  \nonumber    \\
        & \equiv & \;\!\! \wAp\rAp  + \wAm\rAm \,\! = \; 0  \label{prnona}  \\
  \mbox{and}\quad   \rvac &= & \;\; \rAp \quad + \quad \rAm \;\;  = \; 0
   \quad \mbox{for} \quad  \wAp =\wAm
         \nonumber  
\end{eqnarray}   
   This dark sector might then after all be expected to be completely benign, lacking even a net impact upon the classical gravitational field through equations~\ref{Einel}  and \ref{tperf}, and remain utterly undetectable from within the visible matter sector of the Standard Model.

      However, as a further feature for this branch of generalised proper time the structures arising from the symmetry breaking of equation~\ref{spdet} as described for table~\ref{ppdet}
         incorporate the elementary basis for further states of matter. The new fields,  in addition to the gauge fields associated with
       the non-Abelian gauge symmetry  $\slptc$, will couple 
       to the $\Apls(x)$ and $\Amin(x)$ gauge fields.     
      These include 
       a set of $k^2$ Lorentz scalar matter fields based on the $\bdx_{k^2}$ components 
               and denoted $\bphi(x)$ and a set of \mbox{$k$ Weyl} spinor matter fields
          based on the $\bdx_{4k}$
         components
         and denoted  $\bpsi(x)$,  transforming under the internal symmetries that include 
         the  Abelian gauge symmetry $\uo_{\! D}$  in table~\ref{ppdet}, which in turn will be associated with a further 
       gauge field denoted $\Amax(x)$.  
         Since these further matter fields are all presumed to have positive kinetic energy they can be collectively  denoted $\Mplu=\{\Mpls\}$.
       These new states of matter might in principle \textit{also} be created out of the vacuum via interactions with the negative energy $\Amin$ states as depicted in figure~\ref{AAMall}(a).
 

\begin{figure}[htbp]    
\centering
\leavevmode
\includegraphics[width=14.4cm]{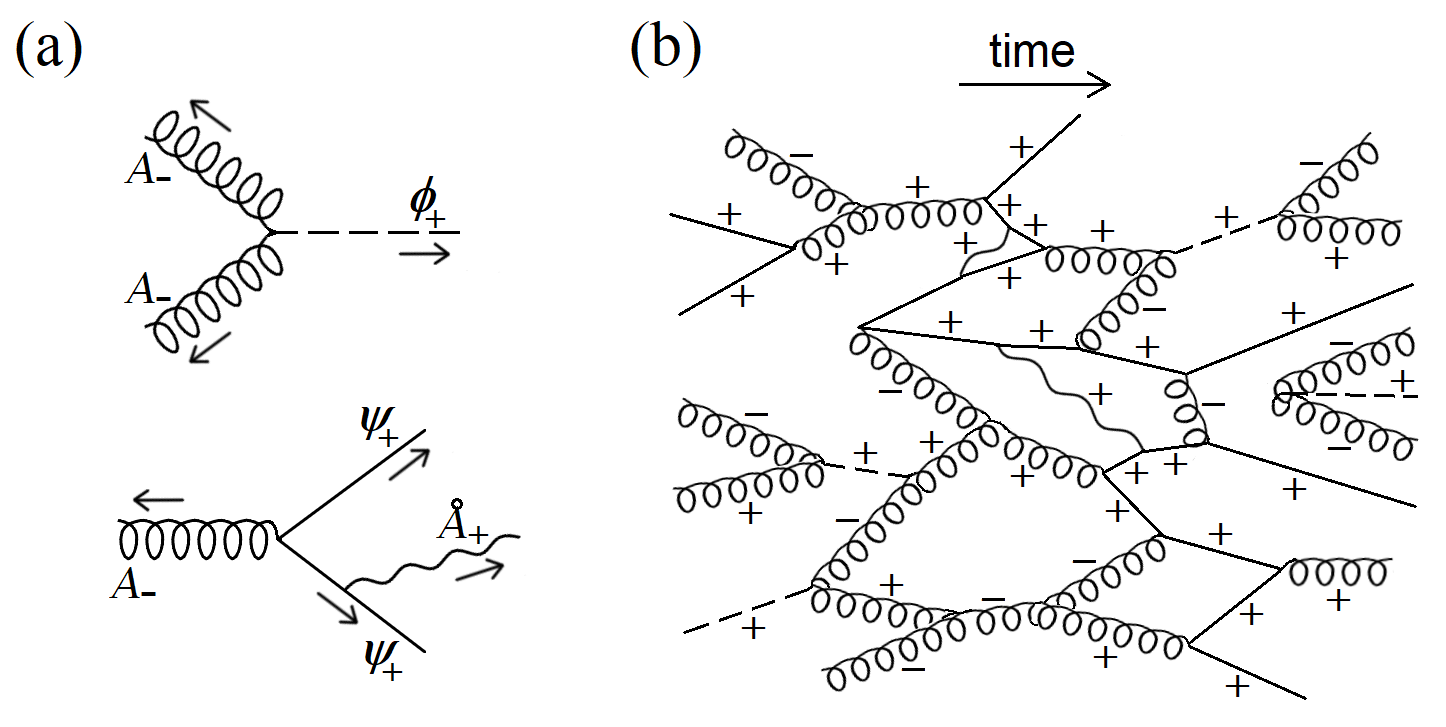}
\caption{\setb 
  The vacuum state generated for a non-compact gauge field in interaction with a source of
  positive energy matter fields.
   (a)   Creation out of the  vacuum of negative energy $\Amin$ gauge bosons as coupled to positive energy 
       $\bphip$   and   $\bpsip,\Amaxp$ matter states  as an additional contribution over figure~\ref{AAonly}(a), again with 
   the arrows indicating forward progression in time and the production of both positive and negative energy states.
    (b) On including these further positive kinetic energy states in this sector
    $\Mplu = \{\bphip \mbox{(dashed lines)}, \bpsip \mbox{(straight lines)},
     \Amaxp \mbox{(wavy lines)}\}$  a network of all permitted interactions,  pictured with  time flowing from left to right for all states and perturbing figure~\ref{AAonly}(b), might again result in a stable equilibrium vacuum state, but now in principle with an asymmetry in the total energy density and pressure balance for this system as described for equations~\ref{prasym} and \ref{eqosV}.
   }
\label{AAMall}
\end{figure}     
 
   The exact nature of the interactions between the $\Amin$ and 
       $\Mplu$ states, such as depicted in figure~\ref{AAMall}(a) and for any further interactions in this sector, 
       will depend upon the implications of the constraints of the theory (deriving from equations~\ref{sbreak} and \ref{gbreak} as will be discussed further for equation~\ref{extgpt}). 
   The possibility and rates of the processes in figure~\ref{AAMall}(a), in themselves or as part of
    higher-order processes involving multiple interactions, will also depend upon  the
     effective couplings and whether the
     $\Apm$ or $\Mplu$ states gain mass.
    With the reverse annihilation processes again possible between the $\Amin$ and $\Mplu$ states, the question will then concern the net properties of a vacuum state  stable equilibrium between the
     $\Amin$ and collective $\{\Apls,\Mplu\}$ states.
     Hence it is the equilibrium vacuum state not for figure~\ref{AAonly}(b) but rather for
     something more like 
      that depicted in  figure~\ref{AAMall}(b) that will generate the empirical phenomena for this dark energy sector.


    The $\Mplu = \{\Mpls\}$ matter component with positive energy density will also presumably make a positive contribution to the pressure, similarly as for the $\Apls$ states of the non-compact gauge field.
     Since this new contribution is \textit{only} on the positive energy and pressure side this inevitably introduces an asymmetry in the dynamics. That is, 
    the effective equation of state parameter $\wAMp$ for the collective positive energy
      $\{\Apls,\Mplu\}$ states  will
       most likely \textit{differ} from the effective equation of state parameter $\wAp=\wAm$ 
      for either the $\Apls$ or $\Amin$ gauge fields alone  described for equations~\ref{pwppm} and \ref{prnona}.
     
      Hence for the stable vacuum state of figure~\ref{AAMall}(b) in general it will \textit{not} be possible to generate both a vanishing total
    pressure  $\pvac=0$ \textit{and} 
         energy density $\rvac=0$  as for
       equations~\ref{prnona}, in place of which we \textit{can} now have:
\begin{eqnarray}
        \pvac &= & \quad \pAMp  \;\;\;\; + \;\;\;\; \pAm  \;\; < \; 0  \nonumber    \\
        & \equiv & \wAMp\rAMp  + \wAm\rAm  < \; 0   \label{prasym}   \\
   \mbox{\underline{and}}\quad   \rvac &= & \quad \rAMp \;\;\;\; + \;\;\;\; \rAm  \;\; > \; 0  
        \quad \mbox{\underline{if}} \quad  \wAMp <\wAm
         \nonumber
\end{eqnarray}       
     That is, for a suitable effective equation of state parameter $\wAMp < \wAm$  a stable vacuum of 
     $\{\Amin,\Apls,\Mplu\}$ states could result with a net excess in \textit{positive} energy density
      $\rvac > 0$,  
  coming from the $\{\Apls,\Mplu\}$ contribution, together with a residual \textit{negative} pressure
   $\pvac<0$ due to the $\Amin$ contribution. Hence the `dark energy' results from the near, but not complete, cancellation of the components of figure~\ref{AAMall}(b), in a manner that would not be possible for the  symmetric $\{\Apls,\Amin\}$-alone vacuum state of figure~\ref{AAonly}(b) and equations~\ref{prnona}.  
    In this way, consistent with equations~\ref{prasym}, a net vacuum equation of state can hence be obtained:
\begin{equation}
   \label{eqosV}
     \pvac= \wvac\rvac \quad \mbox{for} \quad \wvac<0      
\end{equation}    
     with
    negative  $\pvac$ and positive $\rvac$, and both in principle arbitrarily small in magnitude.

     Similarly as for $\wAp$ and $\wAm$ in equations~\ref{prnona},
      here for equations~\ref{prasym} 
    the effective equation of state parameters $\wAMp$ and $\wAm$ may not in themselves have  well-defined
    values, or even meaning, since the associated states are components in a mutually highly interacting system. However, the purpose of these equations is to demonstrate the \textit{asymmetry} between the positive and negative energy components that the $\Mplu$ states introduce. This perturbs the symmetry of the cancellation in equations~\ref{prnona} such that it is now possible to have net $\pvac<0$ together with net
     $\rvac>0$, and even with $\pvac=-\rvac$ and an equation of state parameter $\wvac = -1$ in equation~\ref{eqosV}. 
     In fact $\wvac<0$ in equation~\ref{eqosV}  can be essentially arbitrary and could be compatible not only with the cosmological constant $\Lambda$ case of   $w=-1$ but also with
     the quintessence $-1<w<-\frac{1}{3}$
  or phantom  $w<-1$ scenarios for dark energy, and  could even evolve across the divide between them.

   As the universe expands a corresponding dilution of the overall density of the states
   in figure~\ref{AAMall}(b) 
    would lower the mutual annihilation rate between the negative and positive energy components. However, the creation processes of figures~\ref{AAonly}(a) and \ref{AAMall}(a), with an ever constant production rate, would immediately replenish any such drop in the density of states, in principle maintaining a uniform net energy density $\rvac$ and pressure $\pvac$ for the `vacuum' state through the relations of equations~\ref{prasym}
     and \ref{eqosV}. This construction then implies an essentially `steady state' as the cosmos expands
   with the constant `fuelling' of the   vacuum,
    through creation of real $\{\Amin,\Apls,\Mplu\}$ particles, generating an extremely active high density of states that might also be considered somewhat  `plasma-like'.
        Through the creation process no region of spacetime can be devoid of this `dark vacuum-plasma' material structure.
        
         This uniformity is a key feature required in general for dark energy, in particular if taking a form consistent with a cosmological constant term.
     While this vacuum-plasma consists of a raging sea of real
    $\{\Amin,\Apls,\Mplu\}$ states, undergoing a seething interplay of creation and annihilation, the classical gravitational effect, over many orders of magnitude in scale from the atomic to the galactic, will be calm and benign, with almost all the positive and negative contributions cancelling to leave an arbitrarily low net energy density and pressure.

       A large-scale vacuum-plasma candidate for the dark energy sector,
          as the lowest net energy state profusely active with real $\{\Amin,\Apls,\Mplu\}$ particles, 
           might then be consistent with the Einstein field equation~\ref{Einel} without any  explicit energy-momentum on the right-hand side  and with $\Lambda$ considered a free parameter to be determined, both empirically and in principle theoretically.      
          That is, a vacuum equation for general relativity
            might be realised for this  dark energy 
           candidate of the geometric form:
 \begin{equation}
   \label{Einvac}
       G^{\mu\nu} + \Lambda g^{\mu\nu} = 0
 \end{equation}     
    similarly as for a standard quantum field theory  (QFT) vacuum. 
      However, the nature of the vacuum energy of the standard QFT calculation, which is famously
       estimated to be too large by a factor of $\sim\! 10^{120}$, is very different to that of the dark energy sector considered here. The composition of the former vacuum can be conceived of in terms of  the quantum fluctuations and the fleeting production of \textit{virtual} particle states, all of \textit{positive} energy.  
   The vacuum-plasma state of figure~\ref{AAMall}(b) considered here concerns the 
   production of \textit{real} particle states of both \textit{positive and negative} energy,
    created out of the vacuum through the processes of figures~\ref{AAonly}(a) and \ref{AAMall}(a), the contributions of which to the total energy  largely cancel. Hence there is far less need for any
   `fine-tuning' in order to achieve an exceptionally low energy density.

        Whether  or not the cosmological constant $\Lambda$ case is preferred the question remains, however,
         concerning why such \textit{extraordinarily small} values for the magnitudes of $\pvac$ and $\rvac$ in equations~\ref{prasym} for the vacuum-plasma  should be obtained.
   There are several significant factors that suggest this may be plausible, as consistent with empirical observations and at the level required for
     the $\Lambda$CDM cosmological model.
    A central feature is that this vacuum state can be considered as a \textit{minor perturbation} from the 
     $\pvac = 0$ and $\rvac = 0$ equilibrium described for equations~\ref{prnona}. This concerns the  generalisation from the vacuum structure described for figure~\ref{AAonly}(b) to that described for figure~\ref{AAMall}(b), with the following properties:
 
 \pagebreak
 
\begin{itemize}

    \item{The mutual annihilation between the $\{\Apls,\Mplu\}\leftrightarrow\Amin$ states could be so rapid that each of the effective $\rAMp,\rAm$ contributions  in equations~\ref{prasym}  to the net energy density of the vacuum-plasma state of figure~\ref{AAMall}(b) 
     will itself be much lower in magnitude than the energy density of a Standard Model quark-gluon plasma state that involves only positive energy states (and which is $\sim\! 10^{44}$ times larger than the required dark energy density but sets a useful reference scale). }
 \vspace{2pt}
    
    \item{The production rate of gauge boson states $\Apls$ and $\Amin$, deriving from the non-Abelian gauge theory of the internal symmetry $\slptc$   via the self-couplings of figure~\ref{AAonly}(a) may be much higher than that of the $\Mplu$ states via the processes
     in figure~\ref{AAMall}(a), with for example $\sim\! 2k^2$  $\Apm$ states but only  $k$ $\bpsip$ states and  the $k\to \infty$ limit taken.
      That is, the creation processes depicted in figure~\ref{AAMall}(a), also
      dependent upon coupling and mass factors, may  be highly suppressed.}
\vspace{2pt}
        
     \item{Further, of the processes pictured in figures~\ref{AAonly}(a) and \ref{AAMall}(a) only the quartic production of $\Apls$ and $\Amin$ states may be kinematically permitted for these  tree-level diagrams at first order.
         The  $\Mplu = \{\Mpls\}$  states may then in fact only enter through higher-order diagrams or contribute    in terms of virtual particle states over and above the real $\Apls$ and $\Amin$ particle states.
      The 
    $\Mplu$ states, that introduce the asymmetry, might then only make a very minor contribution.}
 
\end{itemize}      
 \vspace{2pt}

   Hence  the  effective energy  and pressure contribution from the 
    $\Mplu$ states can indeed be extremely small and
    the effective equation of state $\pAMp = \wAMp\rAMp$ 
     for the  $\{\Apls,\Mplu\}$ states in equations~\ref{prasym} may have a parameter value $\wAMp$ very similar
      to $\wAp=\wAm$  
  in the equation of state  $\pApm = \wApm \rApm$ for the  $\Apls$ or $\Amin$ 
    components alone in equations~\ref{pwppm}. The impact would then only be that of a very minor perturbation from the net-zero values
     $\pvac=0$ and $\rvac=0$ for the vacuum state of equations~\ref{prnona}
    to the non-zero values in equations~\ref{prasym} and \ref{eqosV}.
    The high degree of cancellation between the positive and negative energy density and pressure contributions, leaving a small residual dark energy effect, might be considered analogous to the high degree of cancellation between the impact of the positive and negative charges locked in the Earth, leaving a small residual magnetic field effect.

  To sum up, the physical form of matter deriving from the branch of generalised proper time of equation~\ref{spdet} and table~\ref{ppdet} exhibits the collective qualitative properties of darkness, spatial uniformity, $\Lambda$-like temporal constancy, and the means of generating an extremely low positive energy density together with negative pressure; all of which are features desired of any
      proposed physical candidate for the dark energy sector.
   By comparison with the construction of phantom dark energy models that employ a source of negative kinetic energy for a posited scalar field and with the  nature of the vacuum energy calculation in standard QFT,
  the structures arising in the present theory, including a non-compact internal symmetry group and as developed above,
  provide an ideal candidate for the dark energy driving  the observed accelerating expansion
   of the large-scale structure of the universe. 
   Further, as we describe in the following, a connection can also be established with \textit{dark matter} models, as relevant for observations on the galactic scale.
   
   \pagebreak
   
        The non-compact internal $\slptc$ symmetry in table~\ref{ppdet}
        has a total of $2(k^2-1)$ generators, 
        with an equal number of $(k^2-1)$ `non-compact' generators associated with 
         Killing metric diagonal elements $K_{\alpha\alpha}=-1$ and   
         negative energy
          $\Amin$ states together with $(k^2-1)$ `compact' generators associated with $K_{\alpha\alpha}=+1$ elements and  positive energy 
          $\Apls$ states.
         This symmetry further augments the nature of the symmetry between these states described for
           figure~\ref{AAonly} and equations~\ref{prnona}. 
        The 
         $(k^2-1)$ non-compact generators provide a source of negative kinetic energy as well as   negative pressure
         as a key component needed for dark energy. It is the remaining
         $(k^2-1)$ compact generators associated with 
                the maximal compact subgroup $\sukd \subset \slptc$ that provide the
                 potential link with a dark matter candidate.

   As described above the stable dark energy `plasma' of figure~\ref{AAMall}(b) is considered a `vacuum' state in that it corresponds to the lowest energy state in the context of general relativity and the field equation~\ref{Einvac}. Any significant net excess of positive energy and positive pressure states, involving gauge bosons $\Apls$ from the compact subgroup  $\sukd\subset\slptc$ in interaction with  
      $\Mplu = \{\Mpls\}$ states, might then provide a `hidden QCD' contribution and form a `non-vacuum'
       dark matter component in this sector, over and above the dark energy vacuum-plasma state.      
             Incorporating the spinor states $\bpsi$, associated with the $\bdx_{4k}$  components
      and charged under the internal $\uod$ of table~\ref{ppdet}, 
      this `dark QCD' sector would be akin to standard visible quantum chromodynamics (QCD). Such a dark matter candidate 
     might  then be comprised of  `dark glueballs' together 
      with  `dark quarks' 
       potentially confined in `dark hadron'  or  `dark quark nugget'   states for this confining $\sukd$ gauge symmetry,  as analogous to a number of existing models (see for example \cite{Lonsd,BaiL,Gara}).

      The broader dark sector deriving from equation~\ref{spdet} and table~\ref{ppdet} would then incorporate   two 
        different phases composed from the same underlying components, with exchange between the underlying particle  constituents of  these dark matter and dark energy phases  permitted.  
         For example $\Apls$ states in a dark matter glueball could annihilate with $\Amin$ states in the dark energy sea, incorporating a reverse process of figure~\ref{AAonly}(a) and always conserving energy-momentum, leaving the net excess of positive energy in  $\{\Apls,\Mplu\}$ states in the sea that now
     in fact    contribute to the  dark matter.
               With individual dark glueball, hadron or nugget states  seemingly rapidly losing any sense of individual identity the dark matter might still consist of localised  \mbox{4-momentum} conserving `positive energy knots'  propagating like particles  only with less distinct properties or, on the other hand, could act more like  a fluid  permeating across large regions of space.      In either case           
                 gravitating `islands' of this dark matter with a stable overall structure might still form with the requisite empirical properties. A unified dark sector constructed in this way may have some of the features, and explanatory power, of diffusion models~(see for example \cite{Szyd,Beni,Band}).

        In the very early universe the energy density of the dark sector, exceeding that of the visible sector, would be almost uniformly distributed. As the universe expanded and the matter density diluted small variations in the positive energy excess of states could seed the formation of the  extended  gravitational islands  that we recognise today as vast dark matter haloes encompassing visible galactic structures.
         The residual vacuum-plasma state with very low positive energy density, and very low but net negative pressure,  filling the voids between and extending throughout space as described for equations~\ref{prasym} and \ref{eqosV}
                 and potentially consistent with equation~\ref{Einvac}, would then form  the  dark energy component of this sector.

                  That is, the dark energy may be the thinnest possible spread of dark sector material deriving from energy-momentum conserving interactions with the much greater energy density dark matter component. This  origin for dark energy might shed
           further light on the 
          `cosmological constant problem' concerning the extremely low density required for dark energy.
            At around half the present age of the universe the cosmic acceleration associated with this dark energy residual layer could then begin to dominate the cosmic evolution, and prevent further formation of structure on the larger scales, as consistent with observations.
            Both the large-scale cosmic structure and the dark energy background density, consistent with
             equation~\ref{Einvac}, might then essentially `freeze out' during that cosmic epoch.
            
           The  apparent near coincidence between the  cosmological energy densities of 
           the dark matter and visible matter sectors, composing 26\% and 5\% of the total respectively  (\cite{PDG22} section~27), might in a large part be explained for 
              a dark matter sector based on a hidden QCD not unlike the Standard Model QCD that dominates the visible matter sector and as initially generated in parallel.
                Having in turn a  significant interplay between the dark matter and dark energy sectors, with the latter composing 69\% of the total energy density at the present epoch, might then also address the `cosmic coincidence problem' concerning the similarity in the overall dark matter and dark energy density contributions at the current stage of cosmic evolution.

                 That is, this framework could in fact result in a dark energy density, or effective cosmological constant term, that evolves with cosmic time, 
                  providing the possibility of a mutual `tracking' between dark energy and dark matter analogous to that of quintessence models~\cite{Zlat,Tsuji}. This more generally may then encompass the three-way coincidence covering all three sectors of the `cosmological pie chart', with  non-Abelian gauge theories playing a central role in all sectors.

     Despite the negative energy states the dark sector deriving from
     table~\ref{ppdet}, with the dark energy component profusely interacting with the dark matter component, can give rise to stable structures  from our perspective of the visible sector  provided it is   \textit{sealed off} from the Standard Model states 
     associated with table~\ref{esiese},
      other than through the classical gravitational interaction via the curvature of the common
       external 4-dimensional spacetime as essentially fully consistent with general relativity and equation~\ref{Einel}. 
             Any particle-like interactions between the dark and visible sector would imply an instability, via
    a  decay of the vacuum  into
     negative energy $\Amin$ states and Standard Model states of matter.
     
     The `vacuum decay' processes of figures~\ref{AAonly}(a) and \ref{AAMall}(a) are benign since they are contained within the dark sector alone. The quartic diagram in figure~\ref{AAonly}(a) is analogous to processes of vacuum decay considered in 
     phantom models which \textit{are} 
      problematic since they link negative energy phantom states with positive energy Standard Model states,  as  apparently permitted by graviton exchange~(\cite{Cline} figure~1).
      For the present theory the empirical stability of the dark energy vacuum-plasma might itself be considered evidence that gravity is \textit{not} quantised, ruling out such vacuum decay processes involving graviton exchange. 
         In the wider theory the approach to unifying quantum theory and general relativity indeed does \textit{not} involve any `quantisation' of gravity itself~\cite{QGrav}. 
       This `quantum gravity' side of the theory is then fully
         mutually consistent with the dark energy candidate proposed within the overall framework, with
      \textit{no gravitons} to mediate local interactions between the dark and visible sectors as would otherwise imply an instability.

   This  suggests more generally that there are \textit{no}
           particle-like interactions between the dark and visible sectors, unless heavily suppressed, seemingly rendering the possibility of direct laboratory  detection of dark matter states for example highly unlikely.  
           On the other hand, if a branch of generalised proper time could be identified containing a dark matter candidate, but without a dark energy component and negative energy states, there could consistently be a particle-like interaction with the Standard Model, for example through something like a `Higgs portal', without raising the issue of vacuum stability.

       Further contributions to the dark sector may indeed be identified for further possible branches of generalised proper time consistent with equation~\ref{salpha} in augmenting from equation~\ref{seta}.
         For example, equation~\ref{seta} could simply be extended to an \mbox{$(n\!>\!4$)-dimensional} quadratic spacetime form,  as consistent with the special case of equation~\ref{salpha} with $p=2$.
          The resulting non-compact internal symmetry $\sompmm$, with $m_+ + m_- + 4 =n$, 
          provides a basis for a further source of dark energy, with the compact subgroup  $\somdp \subset \sompmm$ as a basis for a dark matter component,  by analogy with the analysis of the branch of
           equation~\ref{spdet} and table~\ref{ppdet}. 
            This would incorporate, at the local level, the particular case of appending $m_+$ extra spatial dimensions,  the generalisation of which led to the present theory.
          
           The key observation is how such dark sectors arise naturally for this approach,
           with a basis in a fundamental theory as constructed 
           upon the generalisation of a  proper time interval, in parallel with the Standard Model sector which is also directly identified as described in the previous section. 
            The fact that there appears to be more than one mathematically permitted dark sector, collectively having a classical gravitational influence on the visible sector but otherwise seemingly undetectable, may also contribute the empirical excess of dark material over ordinary matter. 
   With all sectors deriving directly from the basic structure of the theory, as described for the right-hand side of figure~\ref{contra} and equations~\ref{seta}--\ref{gbreak}, this framework exhibits a 
    high degree of  simplicity and unity as we consider further in the following section.


\section{The Ideal of Unification}
\label{bas5}

   Shortly after the publication of Einstein's theory of general relativity and through the 1920s, alongside the first development of the Kaluza-Klein approach with an extra spatial dimension, several means of augmenting the 4-dimensional spacetime structure to accommodate the electromagnetic field were actually proposed (\cite{Gener} subsection~1.2). Similarly, only now with the need to accommodate a much broader variety of matter, the approach of generalised proper time as described for figure~\ref{contra} and equations~\ref{seta}--\ref{gbreak} provides a further possible means of identifying the elementary structures of matter on
    augmenting the \mbox{4-dimensional} spacetime form.
This new approach to unification involves a change in perspective from a starting point in an extended higher-dimensional spacetime structure to a generalised local form for proper time, hence
 superseding the employment of extra spatial dimensions. This 
 can be seen as a natural  progression from the historical developments  in  the elementary relationships between the most basic entities of space, time and matter (see for example~\cite{susy21} slides 7 and 8).
\pagebreak

          The extraction of the basis for forms of 4-dimensional \textit{spacetime and matter} from forms of \textit{time} alone, as described for equations~\ref{sbreak} and \ref{gbreak}, may sound implausible; however, the aim of deriving essentially all the fundamental structures of physics from the \textit{simplest} possible foundation is precisely the \textit{ideal of unification}. 
        Given that \textit{time} is something that we are intimately familiar with, unlike 
        the hypothetical extra dimensions of
         \textit{space}, this also constitutes a very conservative, as well as simple and unique, starting point for a unified theory.

 The long history of the `least action principle', that plays such a central role in relation to the dynamical equations for particle physics, has also been marked by a central role for time. 
 The original conception that the path of light between two points follows a `principle of least length', dating from
  Heron of Alexandria (circa 40$\,${\small AD}) and as also applies for the reflection of light from a plane surface, was replaced by a `principle of least time'   by Pierre de Fermat (in 1662), 
   incorporating also the refraction of light and more general optical phenomena. 
  Generalising to apply to matter as well as light Joseph-Louis Lagrange (circa 1760) proposed that bodies follow a path that minimises the `action', defined as the time integral of the difference between the kinetic and potential energy. This integrand quantity became known as the `Lagrangian', while the `principle of least action' was later revised to a `principle of extremal, or stationary, action' by William Rowan Hamilton (circa 1835). 
 
   For example,  still within the Newtonian framework, the orbital path of a planet about the sun could be determined by this principle, with the action constructed as an integral over absolute time of a  Lagrangian defined as the difference between the kinetic and gravitational potential energy.
   In general relativity the relativistic version of the action for a planetary orbit in 4-dimensional spacetime simplifies, with 
    any trajectory for a body in a gravitational field described by a geodesic path defined in terms of an extremal duration in proper time. That is, with respect to small variations in the path, a geodesic corresponds to zero change  $\Delta S_{4} = 0$   for an `action' $S_{4}$ defined in
     4-dimensional spacetime based on the extended global metric $g_{\mu\nu}(x)$ from the central part of figure~\ref{contra}:
 \vspace{-3pt}    
 \begin{equation}
     \label{geodic}
     S_{4} \; = \;    \int_{4}  \delta s       \; = \; 
       \int_{4}   \left( g_{\mu\nu} \frac{\delta x^{\mu}}{ \delta s}
                     \frac{\delta x^{\nu}}{ \delta s}  \right)^{\!\frac{1}{2}}  \delta s
 \end{equation}

  In the present theory we are augmenting a local proper time interval from the 4-dimensional spacetime form of equation~\ref{seta} to the generalised form for infinitesimal intervals $\delta s \in \rrr$ over $n$
   components in equation~\ref{salpha}, as pictured on the right-hand side of figure~\ref{contra}. With such intervals invariant under the actions of a symmetry group $\hG$  we might also consider the implications of the invariance, or `stationarity' $\Delta S_{n} = 0$,
   for  finite intervals  or durations   $S_{n}$  in generalised proper time:
 \vspace{-2pt}    
  \begin{equation}
     \label{extgpt}
     S_{n} \; = \;    \int_{n}  \delta s      \; = \;
          \int_{n}   \left( \alpha_{abc\ldots} \frac{\delta x^{a}}{ \delta s}
         \frac{\delta x^{b}}{ \delta s}\frac{\delta x^{c}}{ \delta s}\ldots 
                                                          \right)^{\!\frac{1}{p}}  \delta s
 \end{equation}  
  with the subscripts again indicating the number of components through which $\delta s$ is expressed.
   This form, and the analogy with a Lagrangian formalism, motivates the notation $L_p(\bv_n)_{\hG}$ for the $p^{\mathrm{th}}$-order object in brackets (as employed for example in~\cite{QGrav} equation~28).
    In particular we are interested in the form the integrand in this equation takes under the symmetry breaking described for equations~\ref{sbreak} and \ref{gbreak} as  four preferred components of the  external 4-dimensional spacetime of  equation~\ref{geodic} are locally extracted (see also~\cite{QGrav} equation~43).

     Analysis of the terms that arise for equation~\ref{extgpt} under this symmetry breaking  leads to the identification of apparent `mass terms', with `Higgs' and `Yukawa coupling' factors as alluded to in the discussion following table~\ref{esiese} in section~\ref{bas3}. More generally, constraints can be identified
      from these terms that determine the possible field interactions, as alluded to after figure~\ref{AAMall}
      in section~\ref{bas4} (and expounded in~\cite{QGrav} section~5).
     That some of these terms and constraints could differ from those of the standard Lagrangian picture might, for example, provide a fairly trivial solution for the `strong \textit{CP} problem' (as suggested in the discussion of \cite{QGrav} equation~94).

    The construction of the standard action integral over 4-dimensional spacetime, as employed for both general relativity and the Standard Model, begins with a \mbox{4-dimensional} spacetime arena, a posited set of fields and a proposed Lagrangian, while \textit{postulating} that the extremal action will generate the equations of motion. In the present theory
     we begin simply with time and via the intrinsic arithmetic substructure of equation~\ref{salpha} identify the local ${\delta x^a}$ subcomponents through which the external 4-dimensional spacetime itself together with a matter content can be constructed, as described for equations~\ref{sbreak} and \ref{gbreak}. Since we \textit{begin with time} as an independent entity, variation in the  choice of the derived ${\delta x^a}$ spacetime and matter subcomponents and the gauge field structure cannot change a fixed 
     value of $\int_{n}  \delta s$    but rather \textit{must be} everywhere consistent with the stationarity
     $\Delta S_{n} = 0$ constraint for equation~\ref{extgpt}. Hence, as well as replacing the standard role of a Lagrangian,    this construction can also \textit{provide the basis} for the extremal action principle and \textit{explain why} it works in generating the physical equations of motion.

   The question then concerns the extent of the implications of equation~\ref{extgpt}, under the symmetry breaking of equations~\ref{sbreak} and \ref{gbreak} and the identification of the external  metric $g_{\mu\nu}(x)$ of equation~\ref{geodic} in an extended \mbox{4-dimensional} spacetime, in relation to the Einstein-Hilbert action for general relativity in conjunction with the action in \mbox{4-dimensional} spacetime for the elementary structures of matter. The latter include  the Standard Model, as deriving from the branch of equations~\ref{sedet}--\ref{soctic} and table~\ref{esiese}, as well as the dark sector branches such as described for equation~\ref{spdet} and table~\ref{ppdet}. 
  While underlying the equations of motion for both the spacetime geometry and states of matter, this area of the theory based on
   equation~\ref{extgpt} is also anticipated to shed further light on the means of amalgamating gravity with quantum theory within this framework, as itself a key component in the ambition of unification.


\vspace{-1pt}
\section{Conclusions}
\label{bas6}

   The initial motivation for generalised proper time, as an alternative to the extra spatial dimensions paradigm and as a basis for deriving the elementary structure of matter, has been elucidated with reference to figure~\ref{contra} in
    section~\ref{bas2}.
While the generalised expression for proper time in equation~\ref{salpha}, as employed in equation~\ref{extgpt}, is not unique there are a limited set of possible explicit forms this expression can take as an augmentation from  the local 4-dimensional spacetime form of equation~\ref{seta}, with each branch exhibiting unique mathematical structures. 
           This is the case for the exceptional Lie group branch of equations~\ref{sedet}--\ref{ebreak}, the symmetry breaking of which, as reviewed for table~\ref{esiese} in section~\ref{bas3}, leads to explicit structures of the Standard Model and implications for new physics that might be studied in the laboratory.
           
           Structures of matter deriving from alternative branches of generalised proper time do not have any factor of the  internal gauge symmetry $G$ from equation~\ref{gbreak} in common with the Standard Model branch, and hence will be `dark' in the sense of lacking  gauge  interactions 
      with visible matter (as originally suggested in the discussion of \cite{Gener} equation~42). However, in sharing a common root in the  local substructure of equation~\ref{seta}  and the external 4-dimensional spacetime,  there will be a mutual classical gravitational interaction between all branches through equation~\ref{Einel}, consistent with the cosmological observations of a significant dark sector.

           The branch of  equation~\ref{spdet} and table~\ref{ppdet}, together with others  associated with non-compact gauge groups, can provide the source for dark energy as argued for
         equations~\ref{fym}--\ref{eqosV} and figures~\ref{AAonly} and \ref{AAMall} in section~\ref{bas4}.
          As described there, in addition to being `dark', the physical structure deriving from this branch can exhibit the appropriate properties of near uniformity in space and time together with a very low positive energy density and negative pressure, as key features required for dark energy. 
          It is notable that while it has proven difficult to construct a microphysical model underlying dark energy, here such a candidate arises directly from the elementary structure of the theory.

                  As  described towards the end of section~\ref{bas4}  any positive energy density  excess, as associated with  compact subgroups of the internal symmetry for such dark branches, may accommodate a `hidden QCD' dark matter component, 
      as will tend to gravitate into extended giant haloes.  These structures are then collectively  qualitatively  consistent with  cosmological observations and  the $\Lambda$CDM model in particular.

    The structures of matter deriving from the symmetry breaking of the branches of generalised proper time can then in principle account for all three sectors of the cosmological pie chart. 
    Given that these three sectors of cosmological composition, namely visible matter, dark matter and dark energy,  have very different empirical properties it is striking that they can each have a basis  in the `one simple equation' of the generalised form for proper time in equation~\ref{salpha}.  More generally, as discussed further for equation~\ref{extgpt} in section~\ref{bas5}, the simple basis and broad scope for this approach are very consistent with the traditional goals of unification.
    
     The underlying motivation and simplicity of  deriving  the elementary structures of matter from a local generalisation of proper time as an augmentation from the local  4-dimensional spacetime structure, 
     together with
     the direct manner in which  empirical connections  are simultaneously established  with both the Standard Model and dark sector models, make a compelling case for this approach.
         While the derivation of quantitative empirical predictions will require further development, in this paper the aim has been to show how,
         in comparison with models based upon extra spatial dimensions, 
              the new approach based on generalised proper time  provides a significantly more appropriate foundation for a unified theory.



{\setlength{\baselineskip}{0.87\baselineskip}

\par}



\par}

\end{document}